\documentclass[twocolumn]{aastex631}
\usepackage{mathrsfs}
\usepackage{lineno}
\usepackage{longtable}

\newcommand{\sersic}{S\'{e}rsic}
\newcommand{\hi}{\rm H \textnormal{\textsc{i}}}
\newcommand{\rnine}{{r}_{\rm 90}}
\newcommand{\rfive}{{r}_{\rm 50}}
\newcommand{\ewha}{\rm EW(H\alpha)}
\newcommand{\ewhd}{\rm EW(H\delta_{A})}
\newcommand{\dn}{\rm D_{n}4000}
\newcommand{\mhi}{M_{\rm HI}}
\newcommand{\mstar}{M_{\rm *}}

\newcommand{\Nii}{[\rm N\textsc{ii}]}
\newcommand{\oi}{[\rm O\textsc{i}]}
\newcommand{\oii}{[\rm O\textsc{ii}]}
\newcommand{\sii}{[\rm S\textsc{ii}]}
\newcommand{\nii}{[\rm N\textsc{ii}]}
\newcommand{\oiii}{[\rm O\textsc{iii}]}
\newcommand{\NiiHa}{[\rm N\textsc{ii}]_{6583}/H\alpha}
\newcommand{\OiiiHb}{[\rm O\textsc{iii}]_{5007}/H\beta}
\newcommand{\SiiHa}{[\rm S\textsc{ii}]_{6717,6731}/H\alpha}
\newcommand{\hii}{{\rm H}{\textsc{ii}}}

\defcitealias{Dominguez2018}{DS18}
\defcitealias{Kewley2001}{Ke01}
\defcitealias{Kauffmann2003}{Ka03}
\defcitealias{Schawinski2007}{Sc07}
\defcitealias{Kewley2006}{Ke06}

\shortauthors{}
\graphicspath{{./}{}}

\begin{document}
\title{Exploring the origin of cold gas and star formation in a rare population of strongly bulge-dominated early-type Galaxies}
\author[0000-0001-9472-2052]{Fujia Li}\thanks{E-mail: lifujia@mail.ustc.edu.cn}
\affiliation{Department of Astronomy, University of Science and Technology of China, Hefei 230026, China}
\affiliation{School of Astronomy and Space Science, University of Science and Technology of China, Hefei 230026, China}

\author[0000-0003-1588-9394]{Enci Wang}\thanks{E-mail: ecwang16@ustc.edu.cn}
\affiliation{Department of Astronomy, University of Science and Technology of China, Hefei 230026, China}
\affiliation{School of Astronomy and Space Science, University of Science and Technology of China, Hefei 230026, China}

\author[0000-0001-6083-956X]{Ming Zhu}
\affiliation{National Astronomical Observatories, Chinese Academy of Sciences, Beijing 100101, People's Republic of China}
\affiliation{Guizhou Radio Astronomical Observatory, Guizhou University, Guiyang 550000, People's Republic of China}
\affiliation{CAS Key Laboratory of FAST, National Astronomical Observatories, Chinese Academy of Sciences, Beijing 100101, People's Republic of China}

\author{Ying-jie Peng}
\affiliation{Kavli Institute for Astronomy and Astrophysics, Peking University, Beijing 100871, China}
\affiliation{Department of Astronomy, School of Physics, Peking University, Beijing 100871, People's Republic of China}

\author[0000-0002-6593-8820]{Jing Wang}
\affiliation{Kavli Institute for Astronomy and Astrophysics, Peking University, Beijing 100871, China}

\author[0000-0002-4428-3183]{Chuan-Peng Zhang}
\affiliation{National Astronomical Observatories, Chinese Academy of Sciences, Beijing 100101, People's Republic of China}
\affiliation{Guizhou Radio Astronomical Observatory, Guizhou University, Guiyang 550000, People's Republic of China}
\affiliation{CAS Key Laboratory of FAST, National Astronomical Observatories, Chinese Academy of Sciences, Beijing 100101, People's Republic of China}

\author[0000-0001-8078-3428]{Zesen Lin}
\affiliation{Department of Physics, The Chinese University of Hong Kong, Shatin, N.T., Hong Kong S.A.R., China}

\author[0000-0002-2204-6558]{Yu Rong}
\affiliation{Department of Astronomy, University of Science and Technology of China, Hefei 230026, China}
\affiliation{School of Astronomy and Space Science, University of Science and Technology of China, Hefei 230026, China}

\author[0000-0003-1632-2541]{Hongxin Zhang}
\affiliation{Department of Astronomy, University of Science and Technology of China, Hefei 230026, China}
\affiliation{School of Astronomy and Space Science, University of Science and Technology of China, Hefei 230026, China}

\author[0000-0002-7660-2273]{Xu Kong}\thanks{E-mail: xkong@ustc.edu.cn}
\affiliation{Department of Astronomy, University of Science and Technology of China, Hefei 230026, China}
\affiliation{School of Astronomy and Space Science, University of Science and Technology of China, Hefei 230026, China}

\begin{abstract}
We analyze the properties of a rare population, the strongly bulge-dominated early-type galaxies (referred to as sBDEs) with significant $\hi$ gas, using the databases from the FAST All Sky $\hi$ survey (FASHI) and the Arecibo Legacy Fast ALFA (ALFALFA) survey. 
We select the sBDEs from the Sloan Digital Sky Survey (SDSS) and cross-match with the FASHI-ALFALFA combined $\hi$ sample, resulting in 104 $\hi$-rich sBDEs. 
These sBDEs tend to have extremely high $\hi$ reservoirs, which is rare in previous studies such as ATLAS$\rm ^{3D}$.
70\% of the selected sBDEs are classified as quiescent galaxies, even though they have a large $\hi$ reservoir. 
We study the properties of these sBDEs from five main aspects: stellar population, gas-phase metallicity, stacked $\hi$ spectra, environment, and spatially resolved MaNGA data.
The majority of $\hi$-rich sBDEs appear to show lower gas-phase metallicity and are located in significantly lower-density environments, suggesting an external origin for their $\hi$ gas. 
We find that star-forming sBDEs exhibit statistically higher star formation efficiency and slightly older stellar populations compared to normal star-forming galaxies, suggesting a recent star formation on Gyr-timescale.
They also show narrower and more concentrated $\hi$ profiles compared to control star-forming galaxies, which may explain their higher star formation efficiency. 
\end{abstract}

\keywords{Early-type galaxies; $\hi$ gas; Star formation, galaxy evolution}

\section{Introduction}
Galaxies are typically classified into two broad types: late-type galaxies (LTGs) and early-type galaxies (ETGs), based on their morphology.
ETGs, including ellipticals and lenticulars, are commonly characterized as ``red and dead'' galaxies, owing to a lack of active star formation, and dominated by an old stellar population \citep{Strateva2001,Wetzel2012,Liu2019,Lee2023}. 
The formation and evolution of ETGs are believed to be closely related to the multiple mergers in their lifetime along with the morphological transformation \citep{Croton2006,Park2008,Tamburri2014}, and the radio-mode active galactic nuclei (AGNs) feedback is widely used to maintain the quiescence of star formation in hydrodynamical simulations \citep{Bower2006,Croton2006,Kurinchi-Vendhan2023}.

Most ETGs contain significant quantities of hot gas, but considerably fewer ETGs have been detected with cold gas \citep[e.g., the atomic neutral hydrogen gas $\hi$,][]{Sanders1980,Knapp1985,Sadler2002,Grossi2009,Serra2012}. 
The environment has been found to play an important role in it, as ETGs in low-density environments show a higher cold gas detection rate, and they appear to follow different evolutionary paths compared to the ETGs in clusters \citep{Governato1999,Alighieri2007,Grossi2009,Ashley2019}. 
The ETGs in low-density environments may have experienced a more extended star formation history or rejuvenated star formation activity \citep{Governato1999,Fang2012,Ashley2019}.

There are two avenues for quenched ETGs to obtain cold gas: internal and external mechanisms. 
Internal mechanisms can occur, when evolved stars within the galaxy experience stellar mass loss and this hot material cools down once again \citep{Houck1990,Hofner2018}. 
The primary sources of external gas involve mergers with gas-rich galaxies and accretion of cold gas directly from the environment \citep{Lagos2014,Lagos2015,Davis2016,Griffith2019,Paudel2023}. 
The difference between the two mechanisms lies in the gas-phase metallicity and the dynamics of cold gas. 
The gas from former mechanisms is metal-rich and has almost the coherent dynamics as the stars, while the gas is typically misaligned and metal-poor in external mechanisms, especially in the case of accretion \citep{Davis2019,Bao2022,Lee2023}.
Once they re-acquire gas, ETGs have the potential to trigger star formation again \citep{Fang2012,Lee2023}. 

Ellipticals and lenticulars exhibit significant differences in numerous aspects.
In lenticulars, the presence of the disk indicates that the kinematics are rotationally supported, similar to spiral galaxies, while ellipticals are pressure-supported systems with slower rotation \citep[e.g.,][]{Moran2007,Cortesi2013,Cappellari2016}. 
In this work, we select a rare population, the strongly bulge-dominated ETGs (but more consistent with the distribution of ellipticals in three structural parameters) with $\hi$ detection from two large $\hi$ sample databases, to explore the origin of $\hi$ gas, the basic $\hi$ properties, and the possible triggers for star formation in this rare population.

The Arecibo Legacy Fast ALFA \citep[ALFALFA;][]{Giovanelli2005, Haynes2011} is a blind $\hi$ survey covering $\sim$7000 deg$^2$ of northern sky.
Considering that the $\hi$ detection rate of ETGs is very low, ALFALFA alone could not enable us to obtain a reasonably large sample. 
With the continuous drift scan observations of Five-hundred-meter Aperture Spherical radio Telescope \citep[FAST;][]{Nan2011,Jiang2019,Qian2020}, the FAST All Sky $\hi$ survey \citep[FASHI;][]{ZhangCP2023} aims to obtain more than 100000 $\hi$ sources. 
Therefore, FASHI could complement the ALFALFA survey and potentially provide a substantial quantity of $\hi$ data that corresponds to that of the ALFALFA and even more. 
The large solid angle covered by the FASHI-ALFALFA combined sample provides us the opportunity to statistically study the sample properties of the rare population, and to further explore the origin of cold gas and investigate the triggers for star formation.

The paper is organized as follows. 
We introduce and describe our sample matching and selection from the FASHI, ALFALFA, and SDSS in Section \ref{sec:Sample and data}. 
Section \ref{sec:Sample properties} shows the basic sample properties of our sBDEs sample from the FASHI-ALFALFA combined sample.
Then we present the results of star-forming and quiescent sBDEs separately, including stellar populations, gas-phase metallicity, stacked $\hi$ spectra, environments, and MaNGA observations in Section \ref{sec:Results}.
We discuss the origin of cold gas and the possible triggers of star formation for these $\hi$-rich sBDEs in Section \ref{sec:Discussion}. 
Finally, we briefly summarize our main results in Section \ref{sec:Summary}. 
Throughout this work, we adopt a flat $\Lambda$CDM cosmology with $\Omega_{m}=0.3$, $\Omega_{\lambda}=0.7$ and $H_0=70$ km s$^{-1}$ Mpc$^{-1}$.

\section{Sample and data}\label{sec:Sample and data}
\subsection{ALFALFA and FASHI}
ALFALFA is a blind survey with the Arecibo 305 m radio observatory, covering $\sim$7000 deg$^2$ between 0$^\circ$ and +36$^\circ$ in declination \citep{Haynes2011,Haynes2018}. 
It has a beam size $\sim 4 ^\prime$ and the average rms is about $\sigma_{\rm rms} = 2.0$ mJy at 10 km s$^{-1}$ spectral resolution.

FAST, with a diameter of 500 m and a 19-beam receiver, offers higher sensitivity, deeper observations, and higher observational efficiency in detecting $\hi$ emission lines than ALFALFA. 
FASHI aims to observe an area of $\sim$22000 deg$^2$ between $-$14$^\circ$ and +66$^\circ$ in declination, with the expectation of detecting more than 100000 $\hi$ sources \citep{ZhangCP2023}.
The beam size of FAST is slightly smaller than that of ALFALFA, approximately 2.9$^\prime$ at 1.42 GHz, and the average rms is about $\rm \sigma_{rms} = 1.5$ mJy with a 6.4 km s$^{-1}$ spectral resolution for FASHI.

The parent $\hi$ samples used in this paper are from the 100\% ALFALFA extragalactic $\hi$ source catalog \citep{Haynes2018} and an early version of the first release of FASHI catalog, which contains 31502 and 40343 galaxies, respectively.
It is worth noting that there are some differences between the catalog we're using and the final first release of the FASHI catalog (41741 in the complete first release catalog), but it doesn't change the final results of this work.

\subsection{SDSS}\label{subsec:SDSS}
We select the parent sample from SDSS DR7 with both photometric and spectroscopic measurements, similar to the sample constructed in \citet{Peng2010,Peng2012,Peng2015}. 
The parent sample is obtained directly from the ``PhotoObj View" and ``SpecObj View" in the SDSS CasJobs site\footnote{\url{http://skyserver.sdss.org/casjobs/}}.
We require the galaxies to have clean photometry, clean spectra, and reliable spectroscopic redshift measurements with redshift $z<0.1$, to achieve a complete matching with the combined $\hi$ sample.
We refer to it as the SDSS optical sample, resulting in 266220 galaxies.

We collect stellar mass, star formation rate (SFR), and flux measurements of strong emission lines ($\rm H\alpha$, $\rm H\beta$, $\oii \lambda \lambda$3726, 3729, $\Nii \lambda$6585, $\sii \lambda \lambda$6717, 6731 and $\oiii \lambda$5007) from the MPA-JHU catalog\footnote{\url{https://wwwmpa.mpa-garching.mpg.de/SDSS/DR7/}} \citep{Kauffmann2003,Brinchmann2004}.
All of the emission lines have been corrected for extinction using the \citet{Fitzpatrick1999} Milky Way extinction curve, assuming $\rm R_{V}$=3.1 and the Balmer decrement $\rm H\alpha/H\beta$=2.86 under the case-B recombination \citep{Baker1938}.
Besides, we also retrieve the equivalent width of $\rm H\alpha$ emission ($\ewha$), the Lick index of $\rm H\delta$ absorption ($\ewhd$), and 4000 \AA\, break ($\dn$) from the MPA-JHU catalog, which are typically used to diagnose the recent star formation and star formation history on different timescales \citep[e.g.,][]{Bruzual2003,Kauffmann2003,Le2006,Fumagalli2012,Wang2018,Wang2020}.

We collect three structural parameters: concentration index (C-index, defined as the ratio of $\rfive$ and $\rnine$, that represents the radii enclosing 50\% and 90\% of the flux in $r$-band Petrosian flux), \sersic\ n index, and bulge-to-total mass ratio (B/T).
We obtain $\rfive$ and $\rnine$ from the SDSS catalog directly, and the \sersic\ index n is taken from the NASA Sloan Atlas \citep[NSA; ][]{Blanton2011}. 
We use the bulge-disc decomposition catalog provided by \citet{Meert2015} to obtain B/T in $r$-band with the \sersic +Exponential model.

\subsection{Sample matching and selection}
We use a similar method from \citet{Zhang2019} to cross-match the FASHI-ALFALFA combined sample with SDSS.
We first require the most probable optical counterpart (OC) of each $\hi$ detection to have a spatial separation less than 5$^{\prime \prime}$ with SDSS coordinates and a velocity separation less than 300 km/s. 
Then, to avoid the contamination caused by the large beam sizes of FAST and ALFALFA, we exclude the sources that have multiple SDSS galaxies within 3$^{\prime}$ (for FAST) or 4$^{\prime}$ (for ALFALFA), meanwhile, being within three times velocity width of the $\hi$ line profiles.

We aim to study the properties of strongly bulge-dominated $\hi$-rich galaxies and to understand the origin of their $\hi$ gas. 
We adopt the following method to select our sample:

(i) We cross-match our $\hi$ sample and optical sample with that of the morphological catalog described in \citet[][hereafter \citetalias{Dominguez2018}]{Dominguez2018} to obtain {\tt T-type} for each galaxy. 
In short, the \citetalias{Dominguez2018} morphological catalog is based on Convolutional Neural Networks to classify $\sim$670000 galaxies of SDSS DR7 in Hubble sequence {\tt T-type} and Galaxy Zoo 2 classification scheme.
Furthermore, they train an additional model to classify ellipticals and lenticulars.

(ii) We use morphological {\tt T-type} $\le$ 0 to select ETGs from the total sample and use the additional classification $P_{\rm S0}<0.5$ to further exclude the lenticulars that may be dominated by disk.

(iii) Considering the potential errors in machine learning classification and the limitations of SDSS sensitivity, we apply an additional visual morphological classification using the deeper images from the DESI Legacy Imaging Surveys\footnote{\url{https://www.legacysurvey.org/viewer/}} \citep{Dey2019}. 
In this step, we exclude 56 galaxies that are misclassified, influenced by stars, or have a small apparent axis ratio.

It is worth noting that, although these galaxies are classified as ellipticals in \citetalias{Dominguez2018}, considering their abundant $\hi$ reservoir, we expect them to show some different properties or morphology as usual ellipticals.
In other words, some $\hi$-rich ellipticals in our sample are not entirely smooth, but show some low surface brightness features \citep[may be caused by tidal interactions or mergers, e.g.,][or some newly formed star-forming structure]{George2023}.
Nevertheless, we have still included these galaxies in our sample.
To be precise, we do not refer to it as an elliptical sample but rather as a strongly bulge-dominated ETGs (sBDEs) sample.
Finally, this sBDEs sample consists of 104 galaxies, and Table \ref{tab:data} lists their basic properties.

\begin{figure*}[htbp]
	\centering
	\includegraphics[width=1\linewidth]{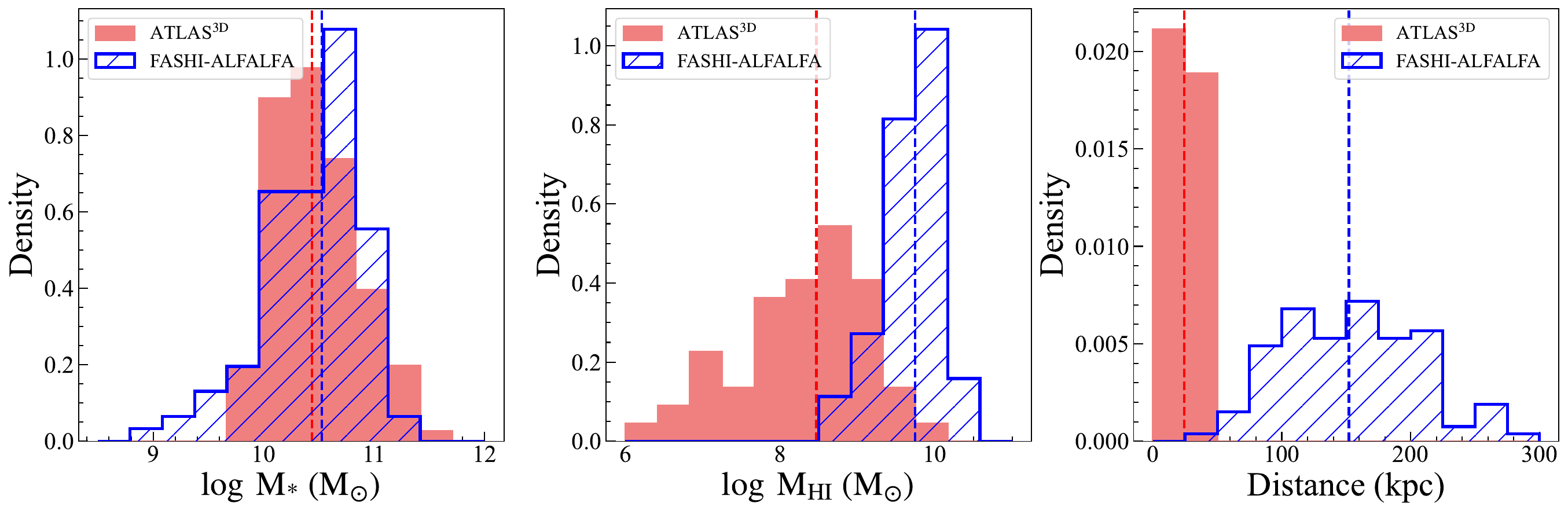}
	\caption{From left to right: histograms of the stellar masses, $\hi$ masses, and distances for the ATLAS$\rm ^{3D}$ (red) and FASHI-ALFALFA combined sample (blue). The median values are indicated with vertical dashed lines. }
	\label{fig:parameter_distribution.pdf}
\end{figure*}

Figrue \ref{fig:parameter_distribution.pdf} shows the distributions of stellar mass, $\hi$ masses, and distances of the ATLAS$\rm ^{3D}$ and FASHI-ALFALFA combined sample.
Compared with the ATLAS$\rm ^{3D}$ survey, which only detected the nearby ETGs within distances less than 42 Mpc, the galaxies in our sample show significantly farther distances (the median value is 152 Mpc). 
Furthermore, ATLAS$\rm ^{3D}$ survey provides a comprehensive and deep $\hi$ observational study of nearby ETGs, while our sample is biased towards $\hi$-rich galaxies with high $\hi$ mass (the median value is $\log \mhi = 9.75$), considering that our galaxies are selected from large flux-limited surveys.
In other words, we focus on a rare population of ETGs that are more bulge-dominated and more $\hi$-rich compared to those in the ATLAS$\rm ^{3D}$.
Such a rare population can only be found within a sufficiently large volume, but the relatively shallow depth of ALFALFA and FASHI may result in completeness issues with our sample.
Specifically, the incompleteness of our sample could impact measurements such as the $\hi$ mass function and the distribution of $\hi$ column densities \citep{Serra2012} if not appropriately corrected. 
Given the limited size of our sample, this study aims to mitigate the influence of sample incompleteness by focusing on comparisons with well-defined control samples. 
The main results in this work are also not sensitive to the effect of sample incompleteness.
Furthermore, we have cross-matched our sample with the NSA catalog to obtain the B/A value of SF-sBDEs and Q-sBDEs and find no differences between them (the median values of 0.872 and 0.877, respectively).
This ensures that the differences in velocity and SFR within SF-sBDEs and Q-sBDEs are not influenced by inclination.

Therefore, the results of this work are still robust. 
It's worth noting that we include both code 1 (high signal-to-noise ratio, S/N) and code 2 (low S/N) objects from ALFALFA. 
This means that the S/N of our sample is not very high, and the median value of S/N is 7.0 from ALFALFA and 14.6 from FASHI. 
This may increase some uncertainty in analyzing the $\hi$ spectra properties.
However, the mass uncertainty, derived following \cite{Jones2018} by combining the uncertainty in the integrated $\hi$ line flux and distance with a minimum of 10\% uncertainty as provided in both the ALFALFA and FASHI catalogs, is not significantly low.
The median value is only 0.06 for both the FASHI and ALFALFA samples.

In addition, we refer to the total $\hi$ sample (after matching with the SDSS optical sample, MPA-JUH catalog, and \citetalias{Dominguez2018} morphological catalog) as the FASHI-ALFALFA combined $\hi$ sample (14518 galaxies, 9468 in ALFALFA and 5050 in FASHI). 

\subsection{$\hi$ profile stacking}\label{subsec:stacking}
We use the stacking technique to study the $\hi$ profile properties of different sub-samples following the method from \citet{Fabello2011}.
For each galaxy, we first shift each $\hi$ spectrum to rest frequency according to the $\hi$ spectral center.
We don't use the central velocity presented in ALFALFA and FASHI release catalogs, which is typically measured as the midpoint between the velocities corresponding to the 50\% of the two peak luminosities.
Instead, we use the flux intensity-weighted central velocity to avoid potential issues caused by the complex $\hi$ spectrum of ETGs when using the above-mentioned method.
Then we stack the $\hi$ spectra weighted by the noise level rms ($w \rm = \sigma^{-2}_{rms}$).
The final stacked spectra are normalized by the integrated luminosity to 1.
It is worth noting that using a different stacking method, such as without weighting, or using a different velocity center in the stacking process does not change our main results.

\subsection{MaNGA observations}\label{subsec:stacking}
We cross-match our final sBDEs sample with galaxies of the Mapping Nearby Galaxies at Apache Point Observatory (MaNGA) survey, which results in six galaxies. The MaNGA survey used integral field spectroscopy to obtain the two-dimensional spectra for more than 100,00 local (0.01 $<$ z $<$ 0.15) galaxies \citep{Bundy2015}. The light is received by two spectrographs of MaNGA, covering the wavelength ranges of 3600-6000$\rm \AA$ and 6000-10300$\rm \AA$, with a resolution of R $\sim$ 2000 \citep{Drory2015}. 

Therefore, the MaNGA view of our six galaxies would provide a glimpse of spatially-resolved properties for the sBDEs sample,  such as stellar velocity ($\rm V_{*}$), H$\rm \alpha$ velocity ($\rm V_{H\alpha}$), H$\rm \alpha$ equivalent width (EW(H$\rm \alpha$)), H$\rm \delta_{A}$ equivalent width (EW(H$\rm \delta_{A}$)) and $\dn$. We adopt the measurements of these properties from the data release of the MaNGA data analysis pipeline \citep[DAP,][]{Westfall2019}. 

\section{Sample properties}\label{sec:Sample properties}
\subsection{$\mstar$-SFR relation}
\begin{figure}[htbp]
	\centering
	\includegraphics[width=1\linewidth]{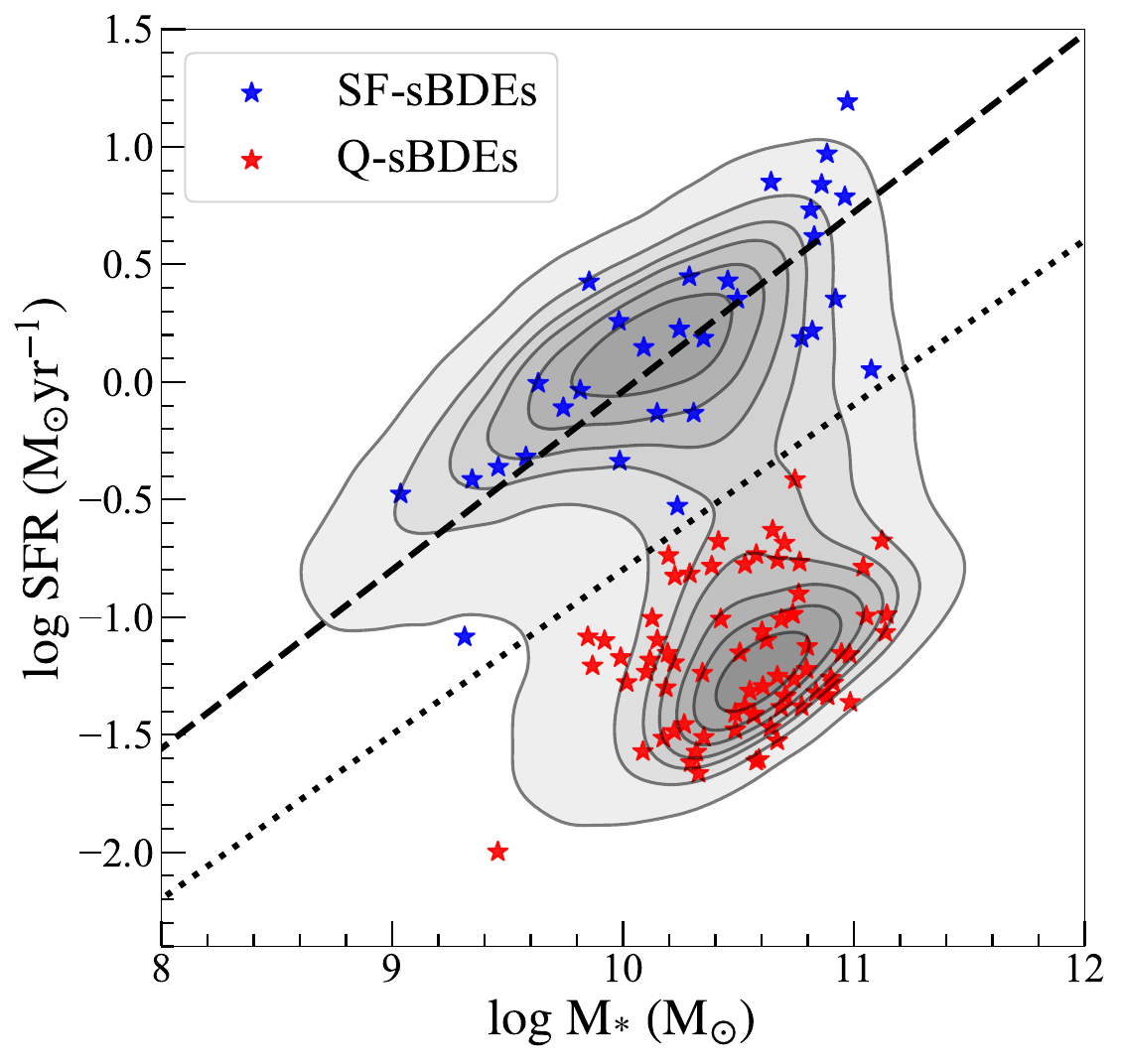}
	\caption{The relationship between SFR and stellar mass for SDSS optical sample (contours) and $\hi$-rich sBDEs (stars). The contours show the distribution of the SDSS optical sample on the diagram, which includes 5\%, 20\%, 35\%, 50\%, 65\%, 80\%, and 95\% of sample galaxies from inside out respectively. The dashed and dotted lines show the star-forming main sequence defined in \citet{Peng2015} and the divide between star-forming and quiescent galaxies from \citet{Zhang2019}. }
	\label{fig:MS_relation.pdf}
\end{figure}

Figure \ref{fig:MS_relation.pdf} shows the SFR-stellar mass diagram from the SDSS optical sample, shown as contours, with our $\hi$-rich sBDEs overlapping on it. 
The dashed line represents the star-forming main sequence defined in \citet{Peng2015} and the dotted line shows the approximate divide between star-forming and quiescent galaxies, taken from \citet{Zhang2019}. 
Star-forming sBDEs (denoted as SF-sBDEs) and quiescent sBDEs (Q-sBDEs) are shown in blue and red color, respectively.

Surprisingly, we find that 31 (30\%) of $\hi$-rich sBDEs are classified as star-forming galaxies, and they almost follow the star formation main sequence, covering a wide range from low to high mass.
\citet{Young2014} has shown that the red sequence contains lots of gas-rich ETGs in the ATLAS$\rm ^{3D}$ sample, as some of them with low mass are in the green valley and even into the blue cloud.
Similarly, our sample also contains some SF-sBDEs but with higher stellar masses ($\log \mstar > 10.7$) and higher SFRs.
The difficulty in forming stars in ETGs may be due to the gas column surface densities being lower than the critical surface density for star formation, resulting in star formation activities occurring in smaller regions and with lower star formation efficiency \citep{Bigiel2008,Oosterloo2010, Serra2012}.
Different from the ATLAS$\rm ^{3D}$ sample, our sample tends to be more $\hi$-rich, making it more likely to trigger star formation over a large region within galaxies, leading to relatively high star formation efficiencies (as later shown in Figure \ref{fig:sample_properties.pdf}).

The SFRs from the MPA-JHU catalog are based on the SDSS spectra which only cover the central 3$^{\prime \prime}$ of galaxies \citep{Brinchmann2004}.
To verify the SFRs from the MPA-JHU catalog, we cross-match our sample with the GALEX-SDSS-WISE Legacy Catalog \citep[GSWLC-A2;][]{Salim2016,Salim2018}, which contains the SFRs derived from broad-band ultraviolet-to-optical photometry with constraints from infrared data.
We find that 88 of 104 galaxies in our sample have counterparts in the GSWLC, and almost all SF-sBDEs are still scattered on the main sequence (only 2 SF-sBDEs are classified as quiescent and 1 Q-sBDE is classified as star-forming galaxies in GSWLC).
The fact that the MPA-JHU catalog does not systematically underestimate the SFR in SF-sBDEs suggests that it seems necessary to trigger central star formation to maintain a high star formation rate. 
In this work, although we use SFRs from the MPA-JHU catalog to classify galaxies into star-forming and quenched populations, the key results do not change when adopting the GSWLC SFRs. 
However, we indeed find more galaxies located in Green Valley using SFRs from GSWLC, and these galaxies may be experiencing low-level star formation at outer regions but not detected by 3$^{\prime \prime}$ diameter SDSS fiber.
We have also checked the GALEX image and found that some Q-sBDEs show weak UV signatures, suggesting that some low-level star formation may also occur in Q-sBDEs.

\begin{figure}[htbp]
	\centering
	\includegraphics[width=1\linewidth]{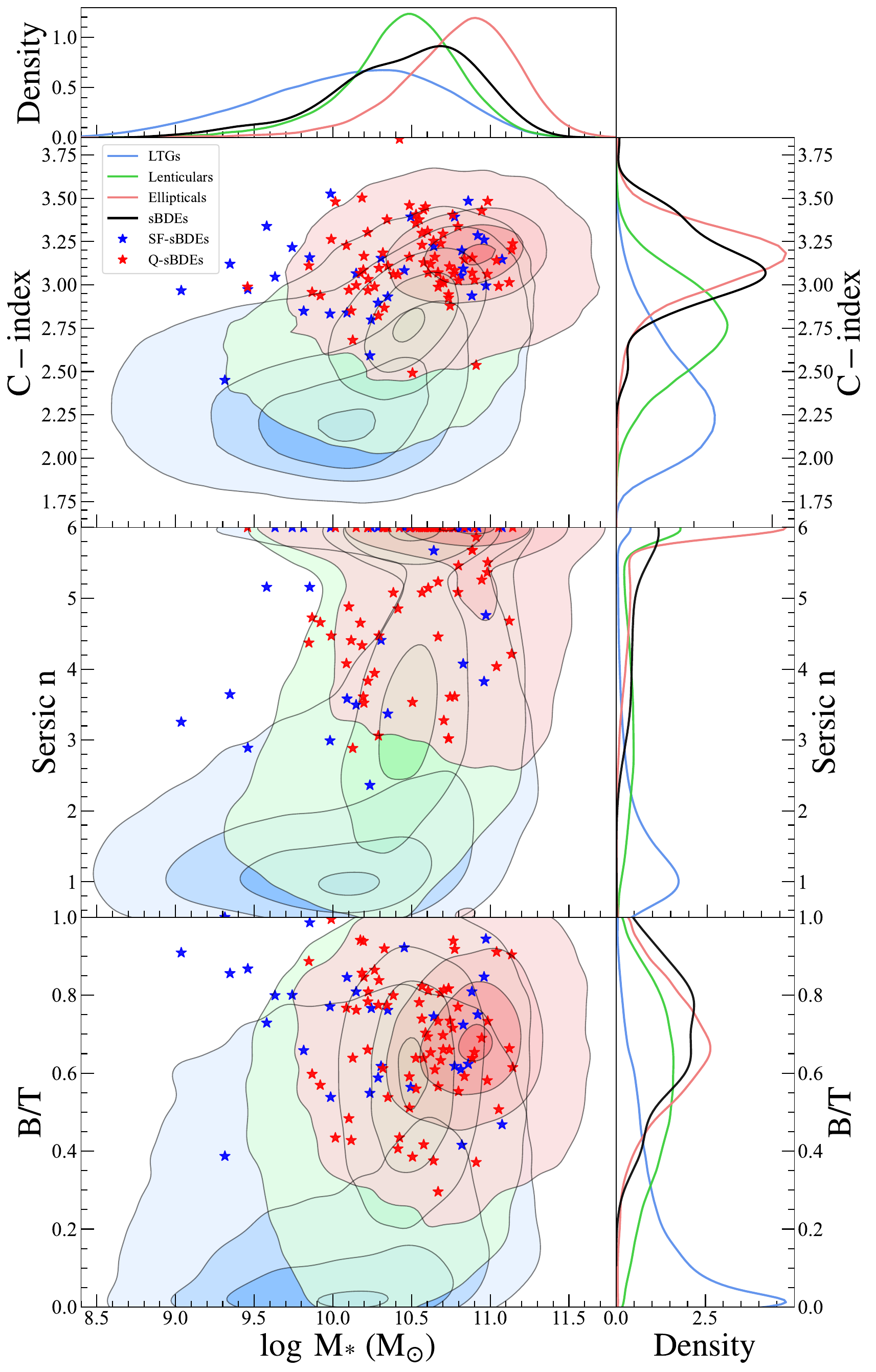}
	\caption{The three subplots, from top to bottom, show the distributions of structural parameters C-index, \sersic\ n index, and B/T as a function of stellar mass. Red, green, and blue represent the ellipticals, lenticulars, and LTGs, respectively, which are classified by \citetalias{Dominguez2018} to our SDSS optical sample. Our sBDEs are in blue (star-forming) and red (quiescent) stars. The four subplots around show the distributions of three samples in terms of their colors, respectively, and the black line shows the distributions of the total sBDEs sample.}
	\label{fig:morphology_test.pdf}
\end{figure}

\begin{figure*}[htbp]
	\centering
	\includegraphics[width=0.95\linewidth]{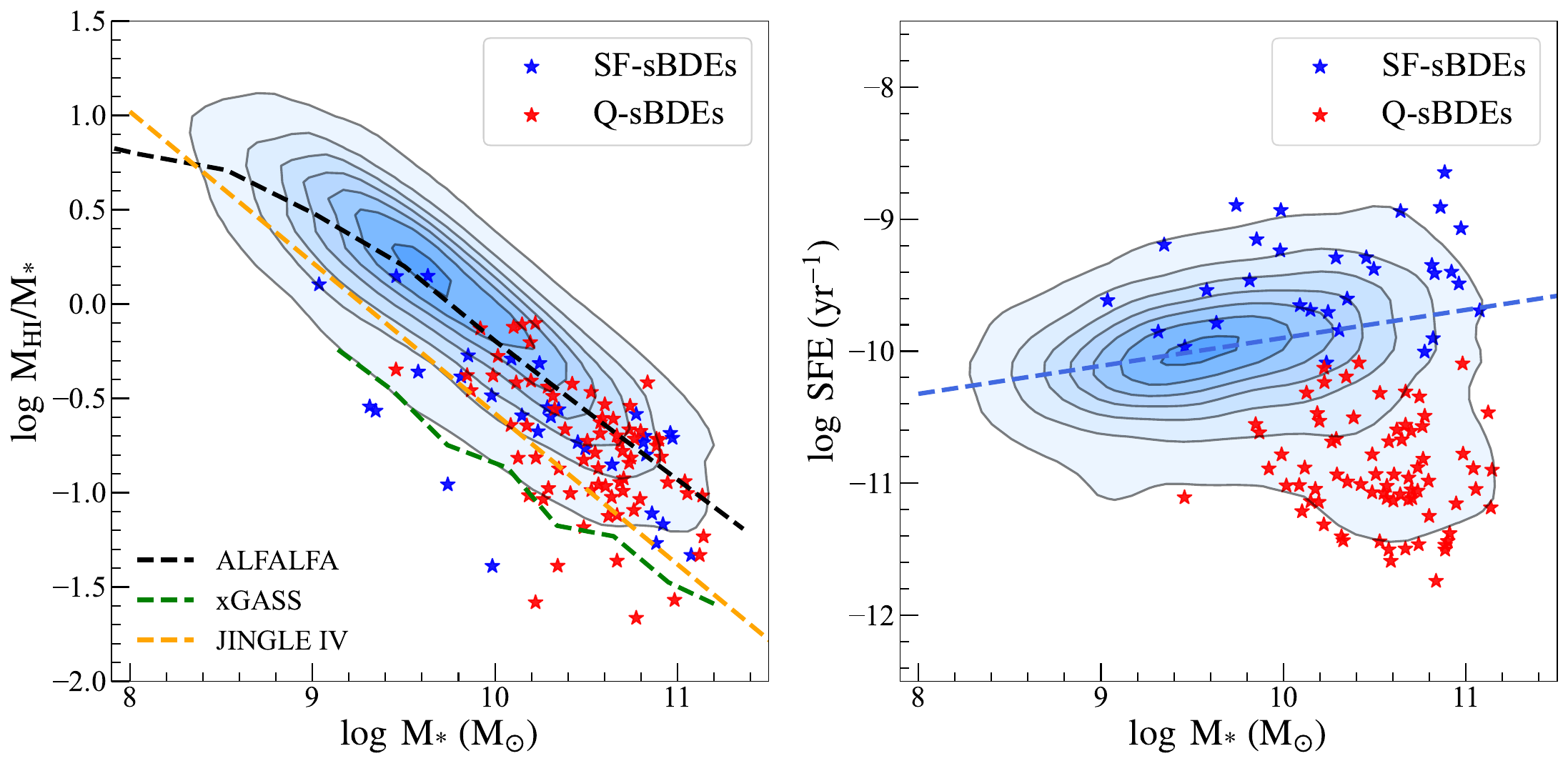}
	\caption{The Correlations of $\hi$ mass fraction ($\mhi/\mstar$, left panel) and $\hi$ star formation efficiency (SFR/$\mhi$, right panel) with stellar mass are shown for FASHI-ALFALFA combined sample (light blue contours), SF-sBDEs (blue stars), and Q-sBDEs (red stars). The black, green, and yellow dashed lines of the left panel show the median relations from ALFALFA, xGASS, and JINGLE IV, respectively. In the right panel, the light blue dashed line shows the best-fitting relation of star-forming galaxies in the FASHI-ALFALFA combined $\hi$ sample.}
	\label{fig:sample_properties.pdf}
\end{figure*}

\subsection{Morphological distributions}

We present the distributions of three different structural parameters mentioned in Section \ref{subsec:SDSS}: C-index, \sersic\ n index, and B/T, shown in Figure \ref{fig:morphology_test.pdf} as functions of stellar mass.
The contours in the left main subplots show the galaxy number density distributions of the ellipticals, lenticulars, and LTGs, which are classified by \citetalias{Dominguez2018} to our SDSS optical sample, with SF-sBDEs and Q-sBDEs overlapping on it.
In the top and right three subplots, we display the distributions of the stellar mass and structural parameters for three samples, each shown in their colors, respectively, and the black line shows the distributions of the total sBDEs sample.

ETGs and LTGs show bimodal distributions in the three structural parameters, with ETGs typically showing higher values of C-index, \sersic\ n index, and B/T \citep[e.g.,][]{Blanton2009}.
Notably, our $\hi$-rich sBDEs are more consistent with the distribution of ellipticals, while some SF-sBDEs also extend to the low-mass end.
We carry out the Kolmogorov-Smirnov (K-S) tests on structural parameter distributions of SF-sBDEs and Q-sBDEs samples.
We find that the p-values between SF-sBDEs and Q-sBDEs are 0.26, 0.32, and 0.54 for C-index, \sersic\ n index, and B/T, respectively, indicating that the hypothesis that these two distributions statistically come from the same population cannot be excluded.
These results validate that these $\hi$-rich sBDEs do share a more similar distribution in structural parameters as ellipticals, although we refer to them as strongly bulge-dominated galaxies to be precise. 

\subsection{$\hi$ and dust properties}\label{subsec:HI properties}

Figure \ref{fig:sample_properties.pdf} shows two correlations associated with $\hi$ gas. 
It is important to note that the contours shown in Figure \ref{fig:sample_properties.pdf} represent the overall FASHI-ALFALFA combined $\hi$ sample (i.e. the $\hi$-detected galaxies), rather than the SDSS optical sample shown in Figure \ref{fig:MS_relation.pdf}.
Therefore, we use light blue and gray contours, respectively, to distinguish them throughout this work.

The left panel shows the relation between $\mhi$ fraction ($\mhi$/$\mstar$) and $\mstar$. 
For reference, the relations from ALFALFA, xGASS \citep[the extended GALEX Arecibo SDSS Survey;][]{Catinella2010,Catinella2018}, and JINGLE IV \citep[][a combined galaxy sample\footnote{Including JINGLE (the JCMT dust and gas In Nearby Galaxies Legacy Exploration), HRS (the {\it Herschel} Reference Survey), KINGFISH (Key Insights on Nearby Galaxies: A Far-Infrared Survey with {\it Herschel}), HAPLESS ({\it Herschel}-ATLAS Phase-1 Limited-Extent Spatial Survey), and HiGH ($\hi$-selected Galaxies in H-ATLAS)}]{DeLooze2020} are overlaid as black, green, and yellow dashed lines, respectively. 
Compared to the FASHI-ALFALFA combined sample (which consists mostly of star-forming disk galaxies), the $\hi$-detected sBDEs generally exhibit less $\hi$ gas. 
xGASS is a mass-unbiased sample with some $\hi$ upper limit sources and JINGLE IV combines several samples with a wide range of properties. 
In comparison to these deeper samples, most of the galaxies in our sBDEs sample have sufficient $\hi$ gas, which can be referred to as $\hi$-rich sBDEs.

The right panel of Figure \ref{fig:sample_properties.pdf} shows the relation between $\hi$ star formation efficiency (SFE$_{\hi}$, defined as SFR/$\mhi$) and $\mstar$. 
The light blue dashed line shows the best fitting of this relation for star-forming galaxies only. 
SF-sBDEs show significantly high SFE$_{\hi}$, with almost all of them lying above the fitting line, suggesting that SF-sBDEs may be undergoing rapid gas consumption and strong star formation activity, while Q-sBDEs exhibit low SFE$_{\hi}$. 

The $\hi$ depletion time, denoted as $t_{\rm dep,\hi}$, is the reciprocal of SFE$_{\hi}$, which is useful to describe the star formation activity and gas consumption \citep[e.g.,][]{WangJ2020,Saintonge2022}.
The median $t_{\rm dep,\hi}$ for the SF-sBDEs sample is 3 Gyr, which is shorter than the typical values of $\sim$9 Gyr for ALFALFA and $\sim$5 Gyr for xGASS.
Nevertheless, the $\hi$ reservoir in $\hi$-rich SF-sBDEs is sufficient to maintain star formation activity for a few Gyrs, as long as it can form $\rm H_{2}$ gas.

\begin{figure}[htbp]
	\centering
	\includegraphics[width=0.95\linewidth]{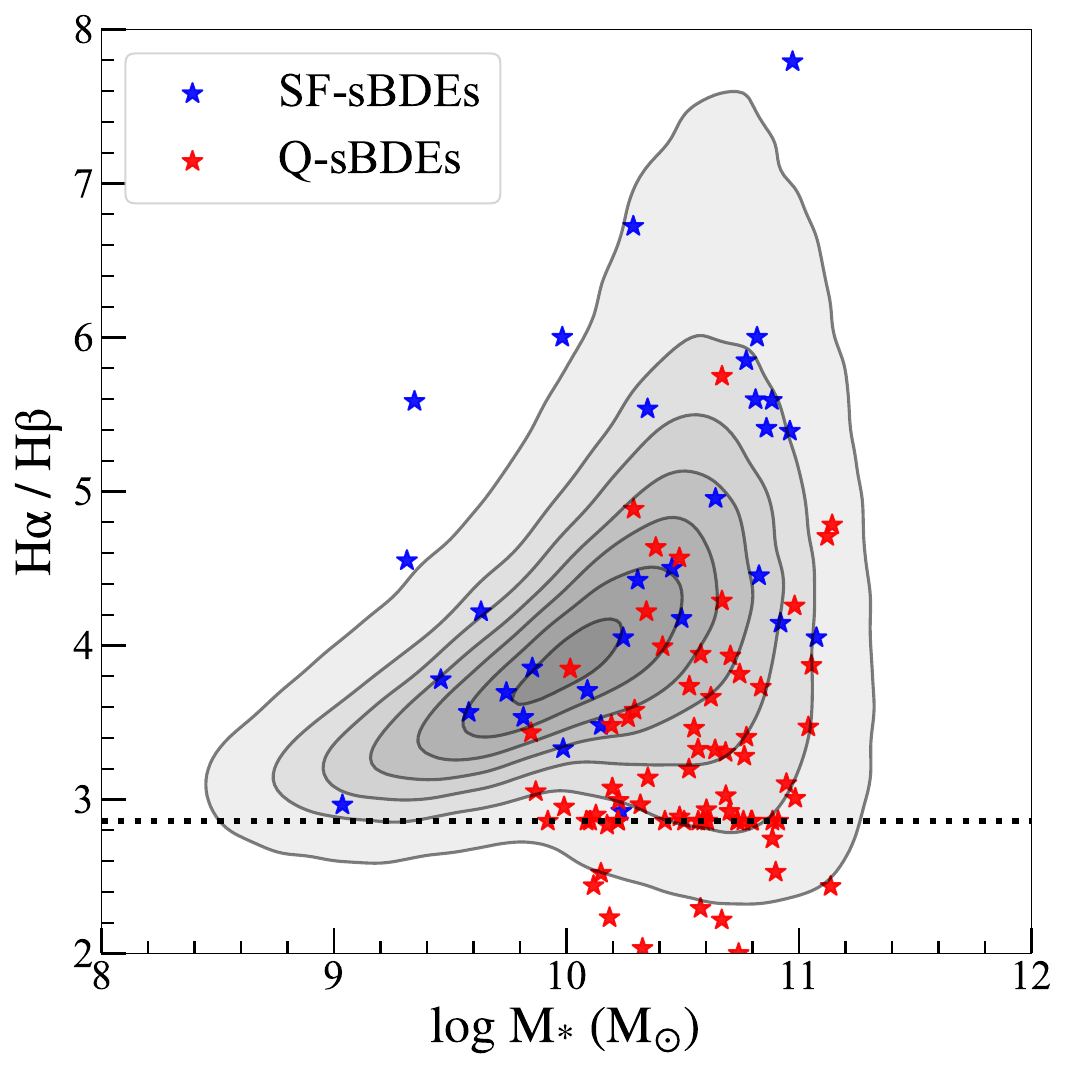}
	\caption{The dust properties estimated by Balmer decrement ($\rm H\alpha/H\beta$) as a function of stellar mass, for SDSS optical sample (contours) and $\hi$-rich sBDEs (stars). The horizontal dotted line shows the intrinsic value of 2.86.}
	\label{fig:extinction.pdf}
\end{figure}

Figure \ref{fig:extinction.pdf} shows the relation between Balmer decrement ($\rm H\alpha/H\beta$, as a proxy for dust content) and stellar mass.
$\rm H\alpha/H\beta$ can be an indicator of dust extinction or dust content, with higher values indicating more dust in galaxies \citep{Kennicutt1992}.  Interestingly, \cite{Barrera-Ballesteros2020} proposed that optical extinction is a reliable proxy for estimating molecular gas content in the absence of direct sub/millimeter observations.  This is consistent with the results in Figure \ref{fig:extinction.pdf} that Q-sBDEs typically have $\rm H\alpha/H\beta$ values around the theoretical value, indicating a lack of dust and molecular gas, while SF-sBDEs have significant dust content and likely molecular gas.

\section{Results}\label{sec:Results}
\subsection{The Stellar Population of $\hi$-rich sBDEs}\label{subsec:Stellar_population}

\begin{figure*}[htbp]
	\centering
	\includegraphics[width=1\linewidth]{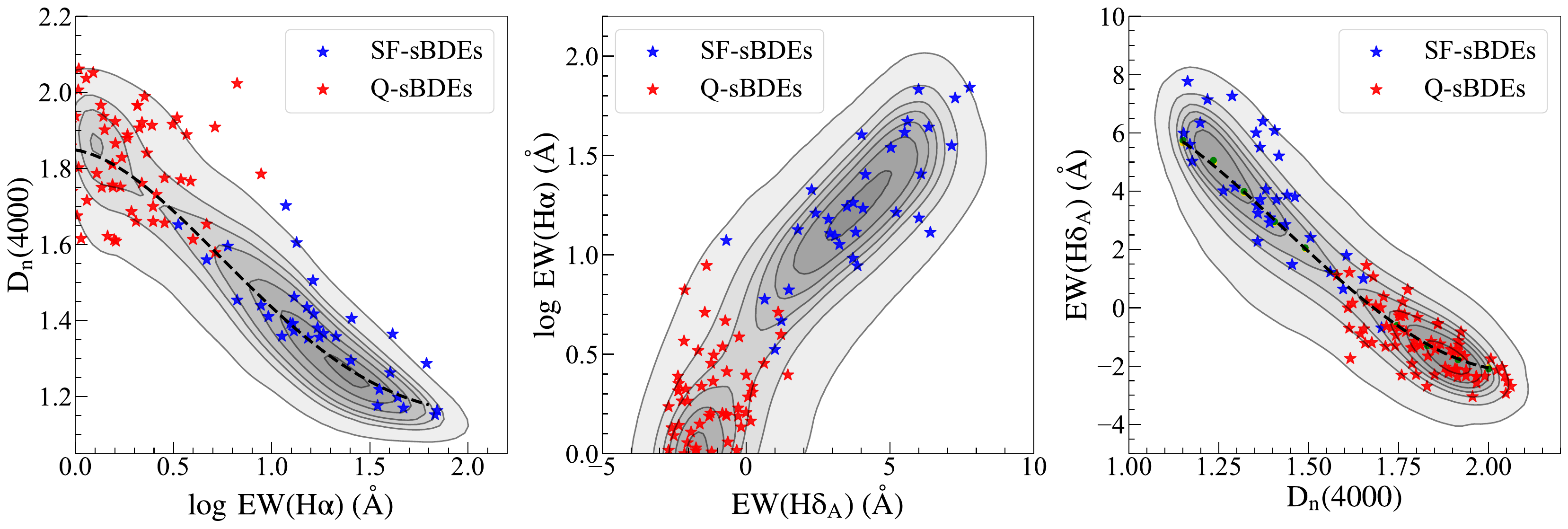}
	\caption{The relations of the $\dn$ with $\ewha$ (left panel), $\ewha$ with $\ewhd$ (medial panel), and $\ewhd$ with $\dn$ (right panel). The dashed lines in the left and right panels show the best polynomial fitting. See the caption of Figure \ref{fig:sample_properties.pdf} for more details on the contours and symbols.}
	\label{fig:stellar_population.pdf}
\end{figure*}

As mentioned in Section \ref{subsec:SDSS}, spectroscopic features such as $\ewha$, $\ewhd$, and $\dn$ are used to diagnose the recent star formation histories on different timescales. With respect to the derived star formation histories from spectral fitting, these parameters are insensitive to dust attenuation and model-independent. 
$\rm H\alpha$ emission traces the recent intense star formation of massive stars within most of their lifetime ($\sim 10$ Myr) \citep[e.g.,][]{Fumagalli2012,Wang2020}. 
$\ewhd$ also traces the recent star formation, but it does so within a longer time scale, up to 1 Gyr \citep[e.g.,][]{Bruzual2003,Le2006}. 
4000 \AA\, break is sensitive to the light-weighted stellar age over the past 2 Gyr, which shows significantly higher values in quiescent galaxies \citep[e.g.,][]{Balogh1999,Kauffmann2003,Wang2018}.
To better understand the star formation properties and star formation history of $\hi$-rich sBDEs, we focus on their stellar populations using three correlations between pairs of $\ewha$, $\ewhd$, and $\dn$, and the results are shown in Figure \ref{fig:stellar_population.pdf}.

It can be seen from Figure \ref{fig:stellar_population.pdf} that most of the SF-sBDEs show young stellar populations and are located in the normal star-forming region of the SDSS optical sample. 
Furthermore, we can find from the left and right panels in Figure \ref{fig:stellar_population.pdf} that SF-sBDEs exhibit slightly larger $\dn$, with a slight departure in distribution.  
This indicates that the stellar populations of SF-sBDEs are slightly older than those of normal star-forming galaxies. 
This is consistent with the picture that old stellar populations dominate these SF-sBDEs, but additionally have a recent star formation activity. 
These young sub-populations of galaxies may have formed from the accreted gas, and it's even possible to discover an inner gas disk and the associated stellar disk component within the galaxies \citep{Oosterloo2010,Young2014}.

Some surveys have demonstrated that UV emission is widespread in ETGs, originating not only from old stars but also from young stars, suggesting the presence of recent star formation in these galaxies \citep[e.g.,][]{Mager2018,Pandey2024}.
The asymmetry and clumpiness of ETGs are significantly larger in the UV than in the optical, and \citet{Serra2012}, based on the ATLAS$\rm ^{3D}$, found that in more than half of the cases, the $\hi$ gas in ETGs is kinematically decoupled from the stars or not disk-like.
Therefore, the relation between $\hi$ gas and stellar populations might be complex.
We will discuss the resolved stellar population of the subset of sBDEs matched with the MaNGA survey in Section \ref{subsec:MaNGA}.

In the literature, some quiescent galaxies could experience recent secondary star formation activity, which is referred to as ``rejuvenation'' \citep[e.g.,][]{Fang2012,Zhang2023,Paudel2023}.
They typically show low-level or medium-level star formation over the most recent $\sim$100 Myr and could transition back to the green valley or even the blue cloud from the red sequence.
We speculate that our SF-sBDEs may experience rejuvenation, which is similar to the scenario of $\rm H_{2}$-rich ATLAS$\rm ^{3D}$ in \citep{Young2014}.
However, these SF-sBDEs show even longer star formation timescales, which can be regarded as strong rejuvenated systems.
Compared to typical rejuvenated galaxies, our SF-sBDEs show relatively large $\ewha$ and $\ewhd$, blending in with normal star-forming galaxy population \citep{Zhang2023}.  
This suggests that these SF-sBDEs may maintain a long timescale ($\rm \sim$1 Gyr or more) rejuvenated star formation, which is consistent with the gas depletion timescales of $\sim$3 Gyr (see the right panel of Figure \ref{fig:sample_properties.pdf}). 

Furthermore, most $\hi$-rich Q-sBDEs overlap with the distribution of normal quiescent galaxies, indicating a lack of recent star formation and the presence of an old stellar population.
This is consistent with the results from ATLAS$\rm^{3D}$ \citep[e.g.,][]{Oosterloo2010,Young2014}.
These galaxies are difficult to trigger medium- or high-level star formation over a long time-scale, whether they accrete gas from the environment or retain gas after they transition to the ETGs.
As a result, they are classified as quenched galaxies and only maintain old stellar populations.
We will further discuss the possible reasons for rejuvenated star formation in Section \ref{subsec:trigger_for_sf}.

\subsection{The BPT Diagrams and Metallicity Properties of $\hi$-rich sBDEs}\label{subsec:Metallicity}
\begin{figure*}[htbp]
	\centering
	\includegraphics[width=0.9\linewidth]{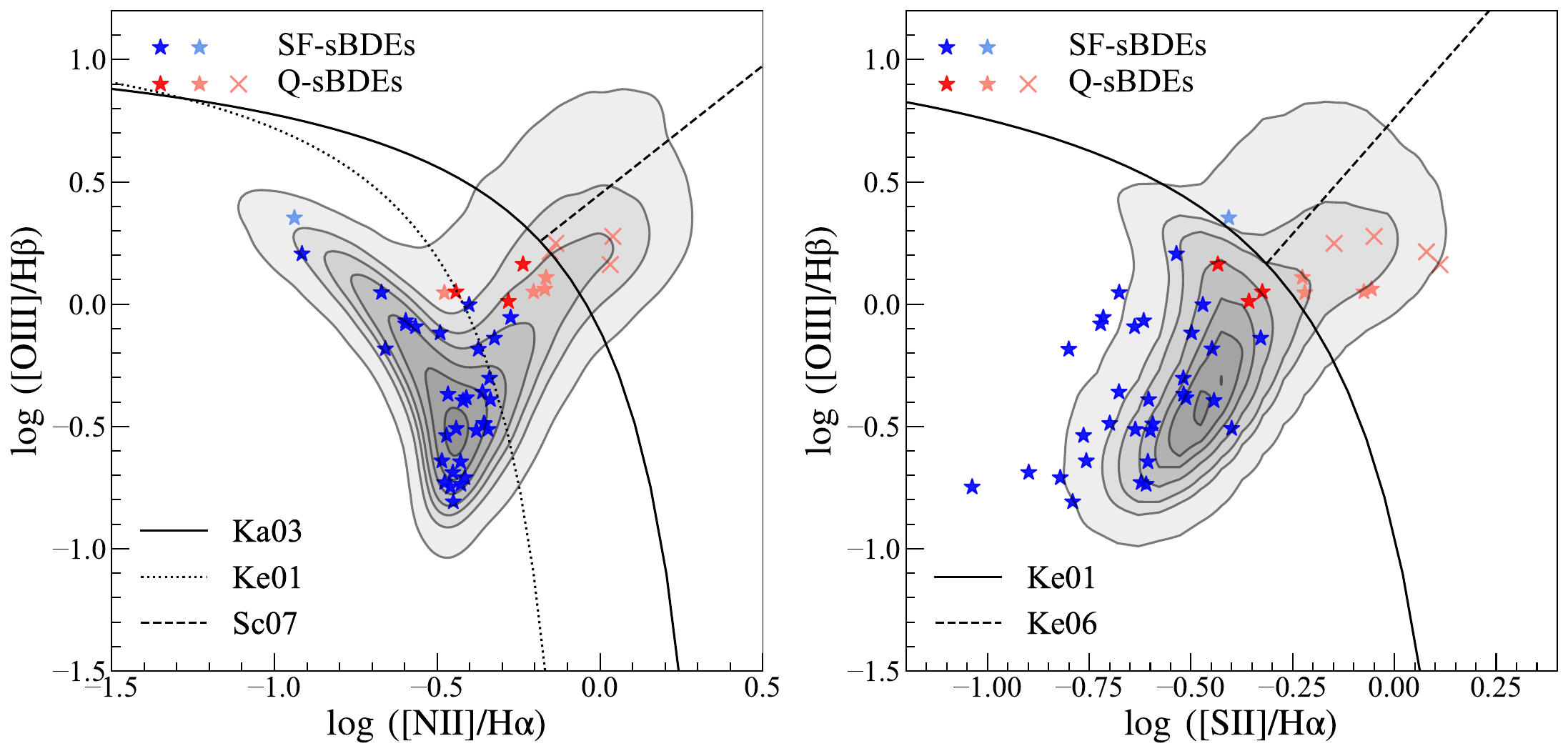}
	\caption{The BPT line diagnostic diagrams for SDSS optical galaxies (contours) and $\hi$-rich sBDEs (stars and crosses) with S/N $>$ 3 in all five emission lines. Left panel: the $\NiiHa$ and $\OiiiHb$ diagnostic diagram. Right panel: the $\SiiHa$ and $\OiiiHb$ diagnostic diagram. The empirical SF/AGN boundary of \citet{Kauffmann2003} (labeled \citetalias{Kauffmann2003}), the theoretical relation of \citet{Kewley2001} (labeled \citetalias{Kewley2001}), and the division lines between Seyfert galaxies and LINER of \citet{Schawinski2007} (left panel, labeled \citetalias{Schawinski2007}) and \citet{Kewley2006} (right panel, labeled \citetalias{Kewley2006}) are shown in dotted, solid, and dashed black lines, respectively. $\hi$-rich sBDEs above the \citetalias{Kewley2001} line in the left panel are shown in crosses and below the \citetalias{Kewley2001} line in the right panel are shown in red and blue stars, respectively. Marked in light red and light blue are those sBDEs below the \citetalias{Kewley2001} line in the $\NiiHa$ diagnostic diagram but above the \citetalias{Kewley2001} line in the $\SiiHa$ diagnostic diagram.}
	\label{fig:BPT.pdf}
\end{figure*}

\begin{figure*}[htbp]
	\centering
	\includegraphics[width=1\linewidth]{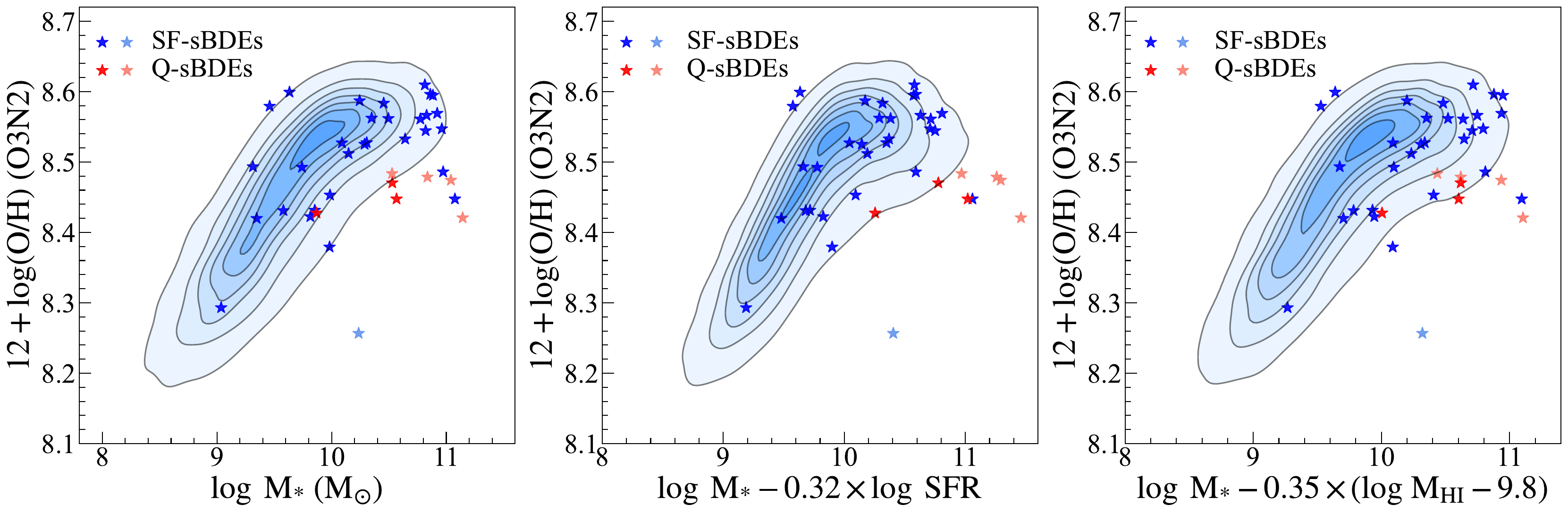}
	\caption{Mass-metallicity relation (left panel), $\rm{FMR_{SFR}}$ using the $\beta$ = 0.32 from \citet{Mannucci2010} (middle panel), and $\rm{FMR_{\hi}}$ using the $\beta$ = 0.35 from \citet{Bothwell2013} (right panel) for the ALFALFA-FASHI combined $\hi$ sample (contours) and $\hi$-rich sBDEs (stars). See the caption of Figure \ref{fig:BPT.pdf} for more details on the color of stars.}
	\label{fig:MZR.pdf}
\end{figure*}

The presence of warm ionized gas in ellipticals does not appear to be uncommon \citep[e.g.,][]{Phillips1986,Falcon-Barroso2006}.
In this section, we aim to study the gas-phase metallicity properties of $\hi$-rich sBDEs, so it's important to select galaxies where ionized gas is dominated by $\hii$ region.

BPT line diagnostic diagram is a useful tool for identifying AGNs and categorizing them into Seyfert galaxies and low-ionization narrow emission-line regions (LINERs) \citep{Kauffmann2003,Kewley2006,Schawinski2007}.
However, some leaked ionizing radiation from $\hii$ regions or old stars can ionize the diffuse gas and form the diffuse ionized gas (DIG), which cannot be explained by the classical model of $\hii$ regions \citep{Kewley2019}.
The DIG can emit some strong low ionization lines, such as $\oi$, $\nii$, $\sii$, and $\oii$, and these emissions can enhance their line ratios with $\rm H\alpha$ \citep{Zhang2017}.
As a result, the DIG can significantly impact the AGNs' diagnostic results from BPT diagrams and may lead to an overestimation of metallicity when using $\NiiHa$ and $\SiiHa$ line ratios \citep{Poetrodjojo2018}.

In order to exclude DIG-dominated galaxies from our sample, the empirical separation of $\ewha$ $>$ 3 \AA\ is useful \citep{CidFernandes2011, Belfiore2016}.  We find that all of the SF-sBDEs and only 15\% of the Q-sBDEs in our sample have $\ewha$ $>$ 3 \AA.
Based on this, we select $\hii$ region dominated galaxies from our sample using two BPT line diagnostic diagrams with $\ewha$ $>$ 3 \AA\, and S/N $>$ 3 in all emission lines we used.
The BPT diagrams are shown in Figure \ref{fig:BPT.pdf}, with our sBDEs shown as stars, overlapping with the SDSS optical sample shown as contours. 

The left panel shows the $\NiiHa$ and $\OiiiHb$ diagnostic diagram, while the right panel displays the $\SiiHa$ and $\OiiiHb$ diagnostic diagram.
Galaxies lying above the lines of \citet{Kewley2001} (solid lines, labeled \citetalias{Kewley2001}) are dominated by AGNs and below the line of \citet{Kauffmann2003} (dotted line, labeled \citetalias{Kewley2001}) are dominated by star-forming galaxies. 
The region between the two lines in the left panel is SF-AGN composites.
The division lines between Seyferts and LINERs are shown as the dashed line from \citet{Schawinski2007} (left panel, labeled \citetalias{Schawinski2007}) and \citet{Kewley2006} (right panel, labeled \citetalias{Kewley2006}).

We use both the two BPT diagnostic diagrams to exclude AGNs based on the following criteria.
We have divided our sample into three categories: (a) the galaxies above the \citetalias{Kewley2001} line in the $\NiiHa$ diagnostic diagram (left panel) are AGNs shown as crosses; (b) the galaxies that lie below the \citetalias{Kewley2001} lines in both diagnostic diagrams are considered as $\hii$-dominated galaxies shown as red and blue stars; (c) those in the composites region in the left panel but above the \citetalias{Kewley2001} line in the $\SiiHa$ diagnostic diagram (right panel) are composite galaxies shown as stars with light colors.

Finally, we have 34 (31 from SF-sBDEs and 3 from Q-sBDEs) sBDEs belong to $\hii$-dominated galaxies and 5 (only 1 from SF-sBDEs and 4 from Q-sBDEs) sBDEs are considered as composite galaxies.
It is worth noting that in our sample, most Q-sBDEs have $\ewha$ $<$ 3 \AA, which is not shown in Figure \ref{fig:BPT.pdf}.

After excluding AGNs from our sample, we derive gas-phase metallicity with the O3N2 calibration from \citet{Marino2013}.
The metallicity of galaxies increases with increasing stellar masses, known as the mass-metallicity relation (MZR), which can serve as an observational constraint for galaxy evolution models \citep[e.g.,][]{Lequeux1979,Andrews2013,Gao2018,Huang2019}.
The MZR is shown in the left panel of Figure \ref{fig:MZR.pdf} and our sBDEs sample is represented as stars, with colors similar to those in Figure \ref{fig:BPT.pdf}.
Although the SF-sBDEs are located on a wide region on MZR, they statistically show lower metallicity, with most of them being below the ridge line of SDSS MZR as indicated by the contours.
Compared to SF-sBDEs, all of the Q-sBDEs have significantly lower metallicity. 

It's worth noting that, even though we have made efforts to exclude galaxies dominated by AGNs or DIG, there may still be contamination in the measurements of gas-phase metallicity of Q-sBDEs.
Considering that DIG can overestimate metallicity with the O3N2 \citep{Kewley2019}, correction of the DIG contamination is expected to strengthen the above result.  The N2O2 metallicity indicator is less contaminated by DIG \citep{Zhang2017}.
Due to the limited number of galaxies meeting the criteria, we present the results in Appendix \ref{appendix:N2O2}.
Using the N2O2 calibration to calculate metallicity does not change our results, and $\hi$-rich sBDEs also show significantly lower metallicity in MZR (N2O2) shown in Figure \ref{fig:MZR_N2O2.pdf}.

We speculate that the cold gas in these sBDEs may come from external mechanisms, i.e., accretion of cold gas from the environment or recent wet mergers with gas-rich galaxies.
Specifically, the lower metallicity of Q-sBDEs is attributed to the recent accretion of cold metal-poor gas from the environment without undergoing metal enrichment. 
As for SF-sBDEs sample, given that most of our SF-sBDEs sample have relatively long star-formation history (see Section \ref{subsec:Stellar_population}), star formation itself can enrich the gas-phase metallicity rapidly \citep{Davis2019}.
This may explain the wide distribution of the SF-sBDEs on the MZR in Figure \ref{fig:MZR.pdf}.
Those sBDEs with low metallicity may have experienced a recent accretion of metal-poor gas, while those with normal metallicity may have already been enriched.
Our findings here are consistent with the metallicity analysis of ATLAS$\rm ^{3D}$ \citep{Young2014,McDermid2015} that the most gas-rich ETGs in low-density environments exhibit low stellar metallicity and signs of gas accretion.

We next focus on studying the fundamental metallicity relations (FMRs) of the sample based on SFR and $\hi$ mass, which can effectively predict the dilution of metal abundances caused by gas inflows and the associated star formation enhancement.
Many works find that MZR shows a secondary dependence on SFR \citep[e.g.,][]{Ellison2008,Mannucci2010,Huang2019} and $\hi$ mass \citep[e.g.,][]{Bothwell2013,Jimmy2015,Chen2022}, which are known as the fundamental metallicity relation ($\rm{FMR_{SFR}}$ and $\rm{FMR_{\hi}}$). 

To better quantify the secondary dependence of the MZR on SFR, \citet{Mannucci2010} introduces a parameter $\mu_{\alpha}$ that combines SFR with stellar mass for the $\rm{FMR_{SFR}}$ relation: 
\begin{eqnarray}
\mu_{\alpha} = \log(\mstar)-\alpha \log(\rm SFR).
\end{eqnarray}
Additionally, \citet{Bothwell2013} defines another parameter $\mu_{\beta}$ to describe the secondary dependence of the MZR on $\hi$ mass:
\begin{eqnarray}
\mu_{\beta} = \log(\mstar)-\beta(\log(\mhi)-9.80).
\end{eqnarray}
We use $\alpha$ = 0.32 from \citet{Mannucci2010} and $\beta$ = 0.35 from \citet{Bothwell2013} in this work to obtain $\rm{FMR_{SFR}}$ and $\rm{FMR_{\hi}}$, respectively.

The final $\rm{FMR_{SFR}}$ and $\rm{FMR_{\hi}}$ relation of ALFALFA-FASHI combined sample (contours) and sBDEs (stars) are shown in the middle and right panel of Figure \ref{fig:MZR.pdf}.
Again, SF-sBDEs show an overall offset in the FMRs, which is more significant in $\rm{FMR_{\hi}}$ with most galaxies located in the low metallicity region.
This supports the above speculation that most $\hi$-rich sBDEs are likely experiencing continuous accretion of metal-poor gas or gas-rich mergers, and the gas inflow rate is larger than the metal production rate \citep{Koppen1999,Gronnow2015}. And this is also true for Q-sBDEs. 

\subsection{The stacked $\hi$ spectra}\label{subsec:Spectra}

\begin{figure*}[htbp]
	\centering
	\includegraphics[width=0.95\linewidth]{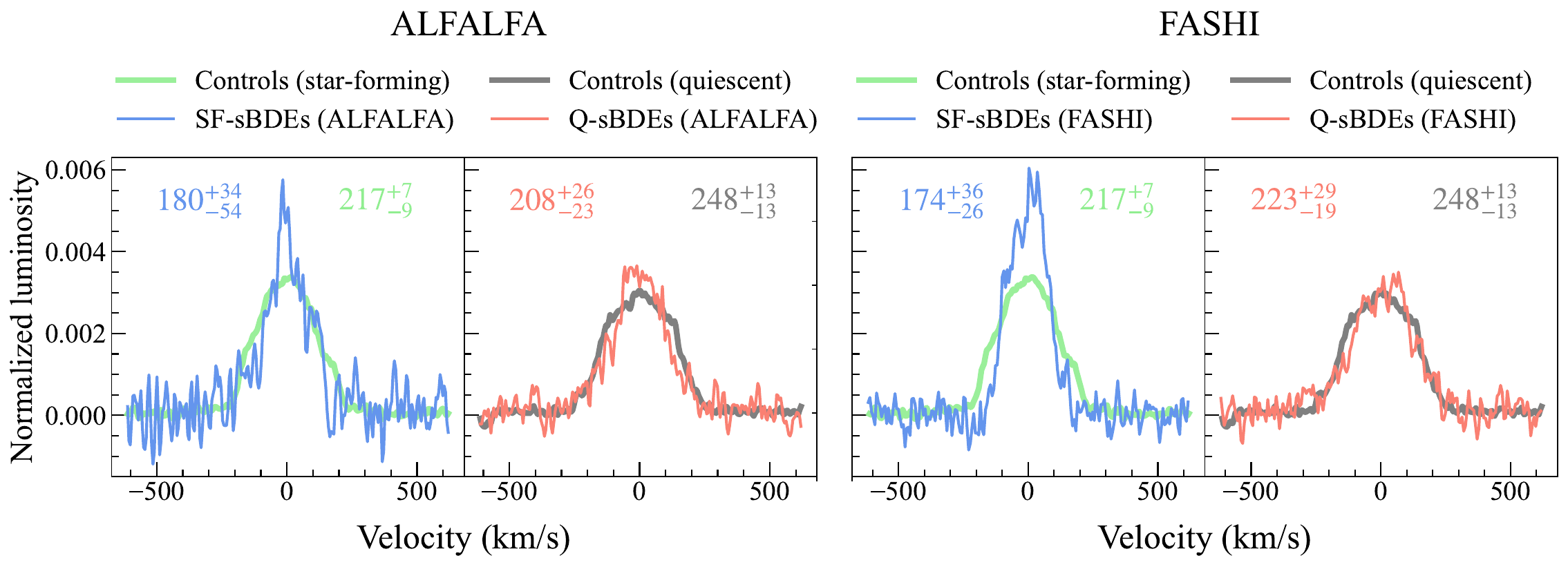}
	\caption{Stacked $\hi$ spectra. Each panel displays the control samples of star-forming and quiescent galaxies from ALFALFA in green and grey colors, respectively, in the background. The left two panels show the stacked spectra for $\hi$-rich sBDEs from ALFALFA, while the right two panels display those from FASHI. The stacked spectra of SF-sBDEs and Q-sBDEs are shown in blue and red, respectively. The stacked spectra are normalized by the integrated luminosity to 1. The velocity width of the $\hi$ profile, $W_{50}$, along with the estimated uncertainty from bootstrapping, is shown in each panel in their respective colors.}
	\label{fig:spectra.pdf}
\end{figure*}

Figure \ref{fig:spectra.pdf} shows the stacked $\hi$ spectra of SF-sBDEs (blue) and Q-sBDEs (red), overlapping with the control samples from ALFALFA.
The reader is referred to Section \ref{subsec:stacking} for more detailed descriptions of $\hi$ profile stacking method. 
The stacked $\hi$ profiles are still quite noisy (especially the SF-sBDEs from ALFALFA) due to the low S/N of some sources.
The control samples are randomly selected from ALFALFA to have the same distribution of stellar mass with SF-sBDEs and Q-sBDEs, shown as green (star-forming galaxies) and gray (quiescent galaxies) colors.
Due to the limited number of quiescent galaxies in the ALFALFA sample (less than 1000 galaxies), we select 500 galaxies for each control sample, allowing for duplicates. 
FASHI has a different rms level and velocity resolution compared to ALFALFA, so we display the stacked $\hi$ spectra from FASHI separately in the right two panels.
However, we overlap them with the same control samples from ALFALFA to facilitate straightforward comparisons.
The velocity widths of the $\hi$ line profile $W_{50}$ are shown in each panel, which are measured at the 50\% level of the peak after smoothing.
The uncertainties of $W_{50}$ are estimated using the bootstrapping method.
For SF-sBDEs and Q-sBDEs sub-samples, we perform 1000 iterations of resampling with the sample size fixed, allowing for repetition.
We also repeatedly select the same number of galaxies for the control samples but from the parent sample for spectral stacking.
The velocity widths and uncertainties of the stacked spectra are adopted by the median values and the values of 16 and 84 percentiles in each panel.

SF-sBDEs appear to have slight differences with narrower, more concentrated, and single-peaked profiles compared to the star-forming control sample. 
The stacked spectra of Q-sBDEs are also slightly narrower compared to the quiescent control sample, but they do not exhibit the concentrated and single-peaked shape seen in SF-sBDEs, instead displaying a flatter profile.
The $W_{50}$ of SF-sBDEs and Q-sBDEs are only marginally larger than control samples, invoking further analysis with larger sample sizes.
Overall, these results highlight the importance of the kinematics of the $\hi$ gas reservoir.
The narrower $\hi$ profile width can be mainly attributed to the slower rise of the rotation curve and more concentrated $\hi$ distribution \citep{Yu2022}.
The wide $\hi$ profile of Q-sBDEs suggests that the gas may be distributed at a larger radius in the outer regions of the galaxy, making it difficult to effectively form stars according to the star formation law \citep[e.g.][]{Bigiel2008, Shi2011}.
These findings are consistent with the results of previous studies that the gas distribution of star-forming ETGs is more centrally concentrated \citep[e.g.,][]{Tacchella2016,Lee2023}.
The shape of the rotation curve may also play an important role. 
However, considering that the $\hi$ distribution and kinematics are likely disturbed and unsettled, and that the information obtained solely from line widths is very limited, high-resolution observations of some of these sBDEs may be necessary in the future.

\subsection{Environments}\label{subsec:Environment}

\begin{figure*}[htbp]
	\centering
	\includegraphics[width=0.9\linewidth]{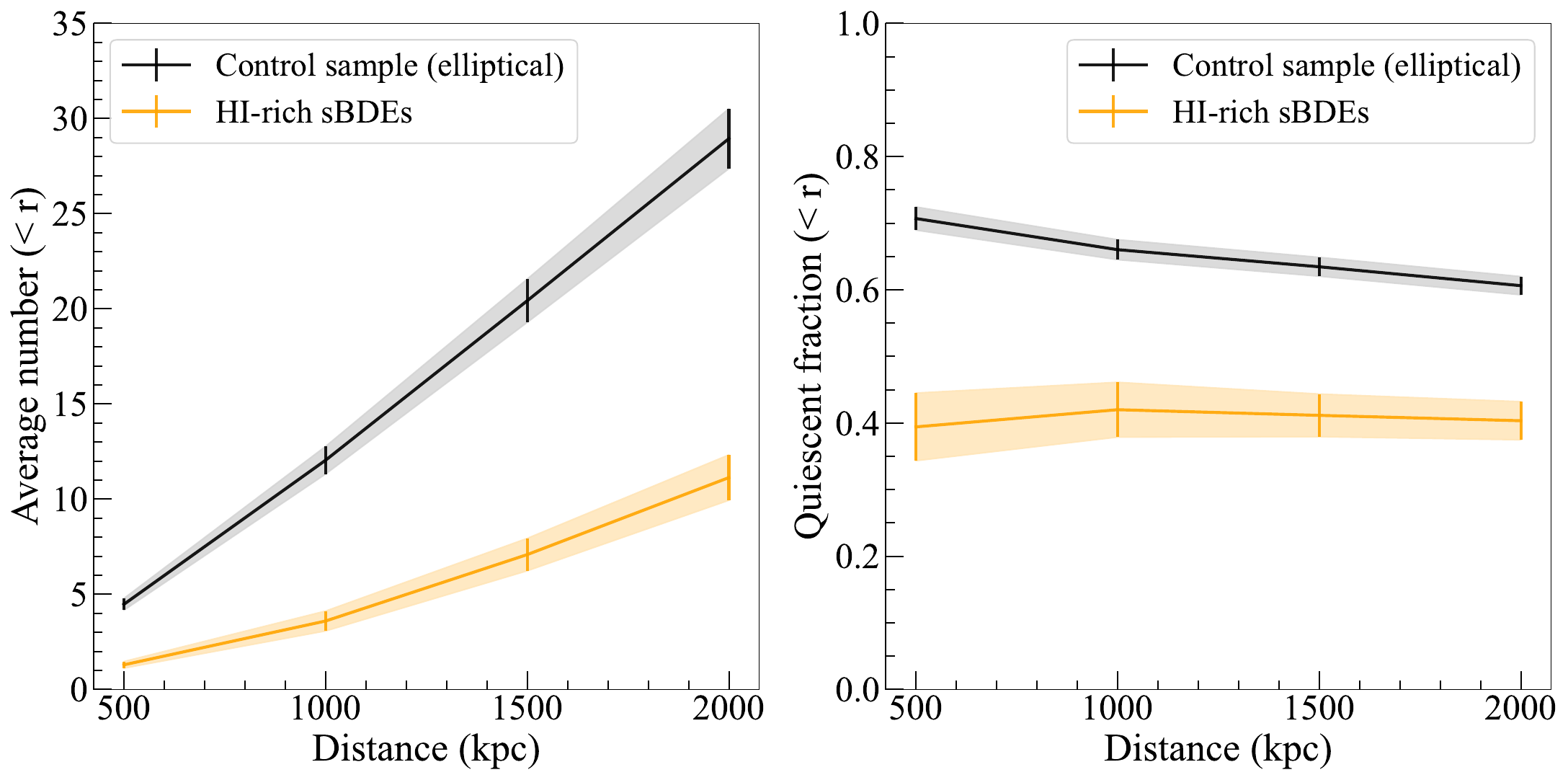}
	\caption{The average galaxy numbers (left panel) and the fraction of quiescent galaxy (right panel) within the radius of 500, 1000, 1500, and 2000 kpc. The orange and black lines show the results for $\hi$-rich sBDEs and the control sample with the same stellar mass and redshift distribution as the total $\hi$-rich sBDEs sample. The uncertainties are estimated using the bootstrapping method.
 }
	\label{fig:environment_1.pdf}
\end{figure*}

\begin{figure*}[htbp]
	\centering
	\includegraphics[width=0.9\linewidth]{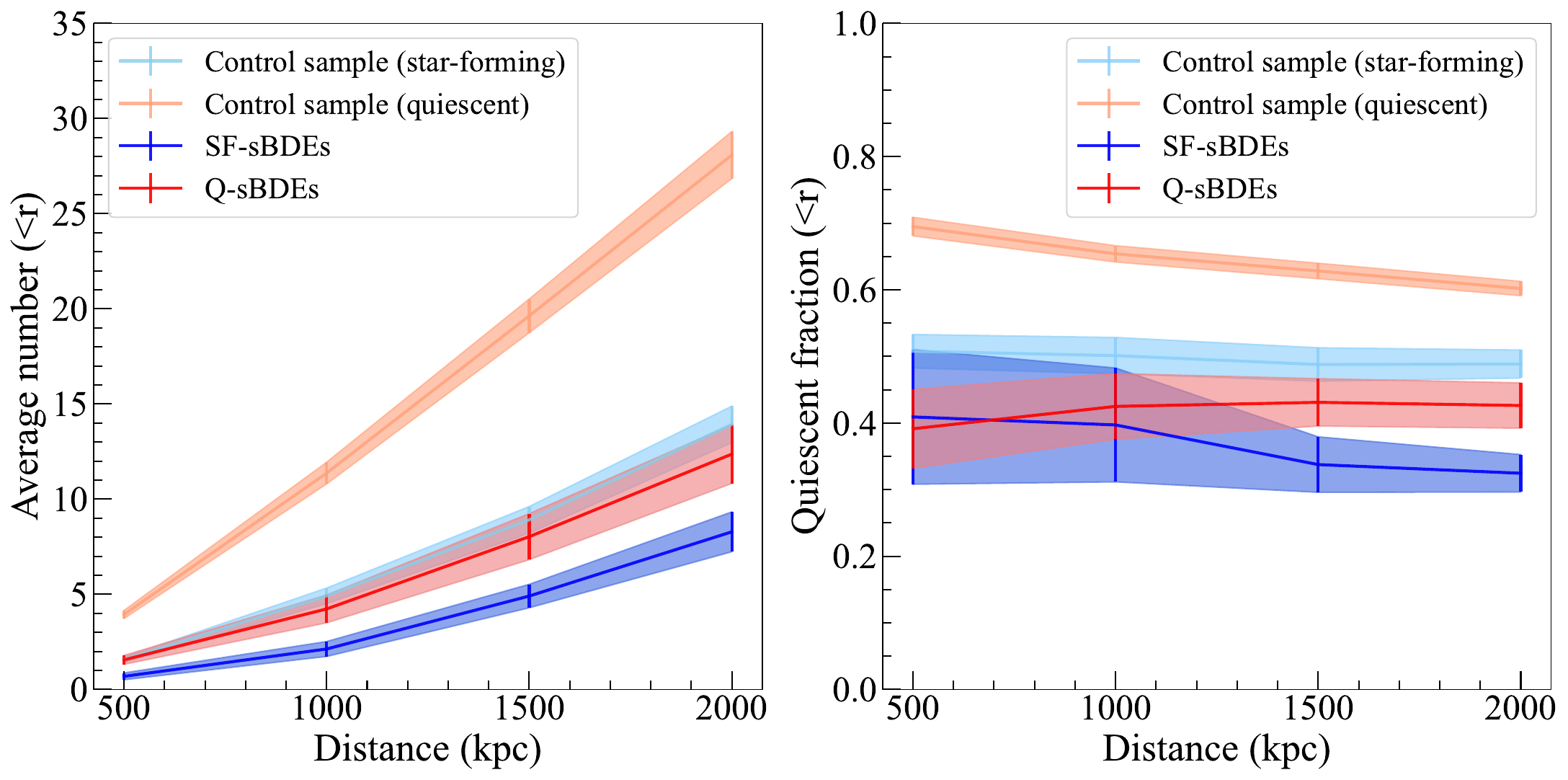}
	\caption{Same as Figure \ref{fig:environment_1.pdf}, but with $\hi$-rich sBDEs dividing into SF-sBDEs (blue) and Q-sBDEs (red). The control samples are constructed separately for SF-sBDEs and Q-sBDEs, which are from the SDSS sample with controlling the same stellar mass and redshift.}
	\label{fig:environment_2.pdf}
\end{figure*}

Many works have found that the $\hi$ detection rate of ETGs and the quenching of star formation strongly depends on the environment density \citep[e.g.,][]{Alighieri2007,Grossi2009,Serra2012}.
In denser environments, such as the central region of galaxy clusters, $\hi$ can be easily stripped from galaxies due to factors like strong tidal stripping, galaxy harassment, and ram-pressure stripping \citep[e.g.,][]{Giovanelli1983,Moore1996,WangJ2021}, leading to a more effective star formation quenching of galaxies \citep[e.g.,][]{Balogh1998,Thomas2010,Lee2023}.

As we have excluded galaxies with multiple objects within 4$^{\prime}$ (for ALFALFA) or 3$^{\prime}$ (for FASHI) and 3 times the $\hi$ velocity widths, we thus investigate the environment of our sBDEs on a larger scale.
For each galaxy, we search from the SDSS optical sample for all galaxies that are located within the velocity difference of 1000 km/s and some certain projected physical distances (500, 1000, 1500, 2000 kpc).
We then count the total number of galaxies and the number of quiescent galaxies using the criteria above for each individual galaxy.
Additionally, we exclude galaxies within 4$^{\prime}$ from these counts to eliminate the bias from our sample selection. 
After completing the counts for each galaxy, we focus on the statistical results of SF-sBDEs and Q-sBDEs samples. 
This provides us with the average number of total galaxies around one galaxy and the overall quiescent fraction at a projected physical distance.

First of all, to understand how the environment affects the gas properties of $\hi$-rich sBDEs, we build an elliptical control sample that is 10 times (1040 galaxies) the size of our sample.
The galaxies in this control sample are randomly selected from SDSS with {\tt T-type} $\le$ 0 and $P_{\rm S0}<0.5$ to match the distribution of stellar mass and redshift with the total $\hi$-rich sBDEs sample. 
Following the same method mentioned above, we calculate the average number of galaxies and quiescent fraction in the environment of the control sample statistically. 

The left panel of Figure \ref{fig:environment_1.pdf} shows the average number and the right panel shows the quiescent fraction for each sample. 
The uncertainties of the average number and quiescent fraction are estimated using the bootstrapping method.
In practice, we perform 1000 iterations of resampling with a sample size fixed, allowing for repetition.
The error bars show the range of 16 to 84 percentiles in both panels.

As shown in Figure \ref{fig:environment_1.pdf}, the normal ellipticals are located in a denser environment and show a higher quiescent fraction of their neighbors compared to $\hi$-rich sBDEs.
This result suggests that the environment plays an important role for $\hi$-rich sBDEs to acquire cold gas.
$\hi$-rich sBDEs potentially reside in lower-density regions and potentially have relatively much more star-forming neighbors on the scale of 1 Mpc. This provides a $\hi$-rich environment, which is the necessary condition for gas accretion from nearby environments.  
Our finding is also consistent with the previous studies that the detection of $\hi$ in ETGs depends on the environment \citep[e.g.,][]{Alighieri2007,Grossi2009,Ashley2019}.
Furthermore, we find that there are only a small number of galaxies within 500 kpc in the sBDEs sample, as shown in the the left panel of Figure \ref{fig:environment_1.pdf}. Consistent with this, most of the sBDEs ($\sim$82\%) are actually central galaxies.  The $\hi$ gas from sBDEs may come from the direct gas accretion from its halo, or the capture of the dwarf galaxies.

Then, we explore the environment of SF-sBDEs and Q-sBDEs, respectively, following the same approach above. 
The control samples of star-forming and quiescent galaxies are constructed separately from the SDSS optical sample, matching stellar mass and redshift with SF-sBDEs and Q-sBDEs, respectively. 
The results are shown in Figure \ref{fig:environment_2.pdf}.
As can be seen, both the SF-sBDEs and Q-sBDEs are more common in lower-density environments and potentially have significantly fewer quiescent neighbors than the quiescent control sample, but they are almost similar to the star-forming control sample.
In comparison to Q-sBDEs, SF-sBDEs only show slightly lower average galaxy numbers in the environment.
For the quiescent fraction in the right panel of Figure \ref{fig:environment_2.pdf}, SF-sBDEs do not exhibit significant differences compared to Q-sBDEs within 1 Mpc, only showing a slightly lower quiescent fraction at larger projected physical distances.

These results are consistent with the findings of ATLAS$\rm ^{3D}$ surveys that the $\hi$ properties of ETGs strongly depend on environment density \citep{Serra2012}.
Based on the FASHI-ALFALFA sample, we find that $\hi$-rich sBDEs are not only more common in lower-density environments but also potentially have significantly fewer quiescent neighbors than the elliptical control sample and the star-forming control sample from SDSS.
On the other hand, ATLAS$\rm ^{3D}$ sample was also observed CO J=1-0 and J=2-1 emission to study the molecular gas properties of ETGs \citep{Young2011,Young2014}.
They find that the environment plays only a weak role in CO detection of gas-rich ETGs and the galaxies within the cluster have the potential to retain molecular gas.
These retained molecular gas could fuel star formation in galaxies, which is why SF-sBDEs and Q-sBDEs samples do not show significant differences in the average number and quiescent fraction in the environment (Figure \ref{fig:environment_2.pdf}).

\subsection{MaNGA view}\label{subsec:MaNGA}
\begin{figure*}[htbp]
	\centering
	\includegraphics[width=0.95\linewidth]{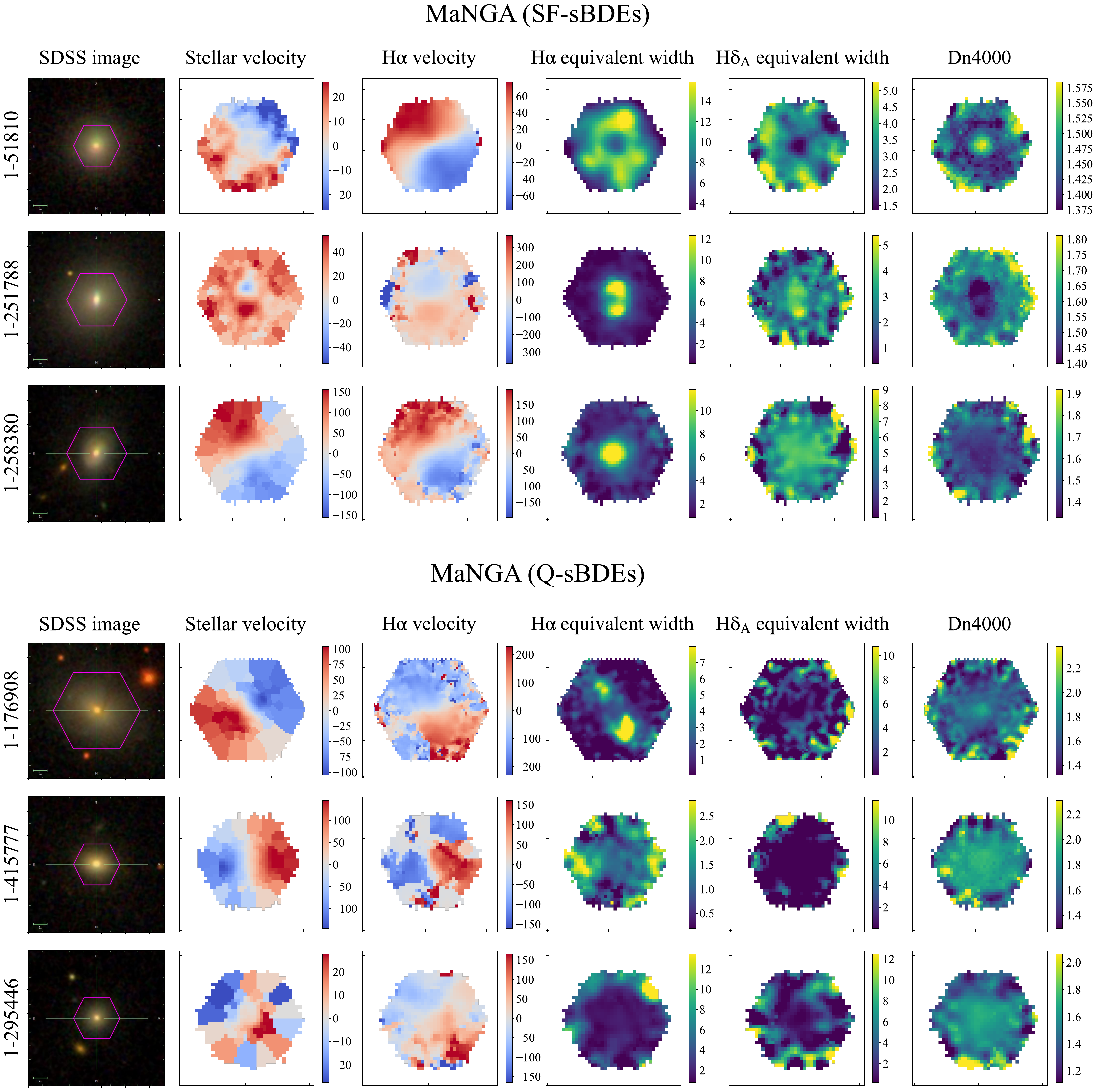}
	\caption{The three SF-sBDEs (top three rows) and three Q-sBDEs (bottom three rows) observed in MaNGA with the corresponding MaNGA-ID marked in the left. Each column shows, from left to right: 1. SDSS g-r-i image with the MaNGA hexagonal Fov overlaid; 2. the stellar velocity fields; 3. H$\rm \alpha$ velocity fields; 4. H$\rm \alpha$ equivalent width; 5. H$\rm \delta_{A}$ equivalent width; 6. $\dn$. These maps are obtained from the MaNGA DAP.
 }
	\label{fig:MaNGA_new.pdf}
\end{figure*}

In Figure \ref{fig:MaNGA_new.pdf} we display the spatially resolved properties of 3 SF-sBDEs (top three rows) and 3 Q-sBDEs (bottom three rows) that are also in the MaNGA survey. 
From left to right, each column shows the SDSS color image, $\rm V_{*}$, $\rm V_{H\alpha}$, EW(H$\alpha$), EW(H$\delta_{A}$), and $\dn$, respectively. 
Although there are only 6 galaxies from our sample in MaNGA, they show diverse properties, allowing us to better understand how these sBDEs obtain gas.

First of all, we have identified two significantly gas-versus-stars misaligned galaxies (MaNGA-ID: 1-51810 and 1-176908) and two slightly misaligned galaxies (MaNGA-ID: 1-258380 and 1-415777). 
The kinematic misalignments of these galaxies suggest that they have experienced external gas accretion or gas-rich misaligned mergers.
Such a high proportion of misaligned galaxies (4/6) indicates that our sBDEs are good candidates for misaligned galaxies, deserving further observations in the future.

Furthermore, we also find two SF-sBDEs show a rotation feature in the central region of galaxies (1-251788 and 1-258380). Especially MaNGA 1-251788, shows a small rotating structure at inner regions with respect to a non-rotating stellar velocity field at large radii. The inner regions show an enhanced specific SFR (high EW(H$\alpha$)) and old stellar population (high EW(H$\delta_{A}$) and low $\dn$).  The misaligned Q-sBDE (MaNGA 1-176908) is also forming stars likely in a newly formed disk, which can be seen in the panel of EW(H$\alpha$).
On the other hand, the Q-sBDE, MaNGA 1-295446, does not show significant recent star formation indicated by H$\alpha$ emission, while EW(H$\delta_{A}$) map shows that this galaxy has a wide area of star formation in last Gyr timescale. 

Overall, these sBDEs show many special properties, indicating that the acquisition of cold gas can result in significant differences in galaxy properties, even for Q-sBDEs. These recently obtained cold gas appear to form a gas disk inside the original stellar components. If this is true, the growth of the disk would potentially modify the galaxy morphology. This is also the reason that we do not refer to these sBDEs as ellipticals, even though they are classified as ellipticals from the machine learning sample. We refer to the event in ellipticals as ``morphology rejuvenation'', where the accreted cold gas forms stars on a disk and ultimately changes the galaxy morphology. We plan to examine whether this is the case in the future. 

\begin{figure*}[htbp]
	\centering
	\includegraphics[width=0.95\linewidth]{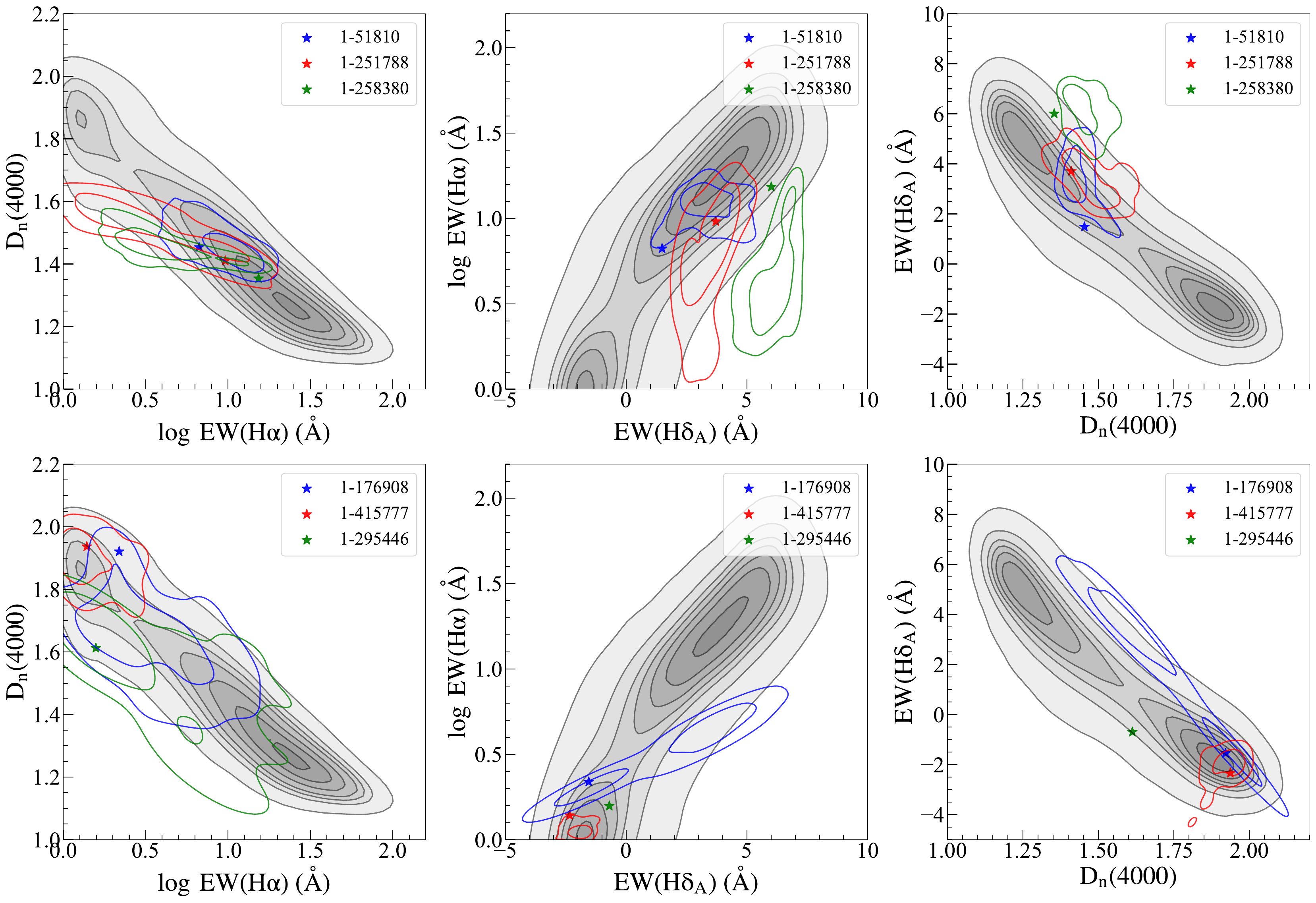}
	\caption{The spatial stellar population properties of three SF-sBDEs (top panels) and three Q-sBDEs (bottom panels) matched to MaNGA, overlapped on the relations of the $\dn$ with $\ewha$ (left panel), $\ewha$ with $\ewhd$ (medial panel), and $\ewhd$ with $\dn$ (right panel). The stars show the measurements from SDSS spectra and the contours show the spatial measurements from MaNGA in corresponding colors. 
 }
	\label{fig:stellar_population_MaNGA.pdf}
\end{figure*}

To compare the differences between SDSS spectra and the spatial stellar population properties of galaxies, we tested the distribution of this subset of MaNGA data in stellar population relations.
We get the $\dn$, $\ewha$, and $\ewhd$ at each point with S/N larger than 5 within the MaNGA field for these galaxies.
Figure \ref{fig:stellar_population_MaNGA.pdf} shows the comparison of the three SF-sBDEs (top panels) and three Q-sBDEs (bottom panels) galaxies measured in MaNGA and SDSS.
The stars show the measurements from SDSS spectra and the contours show the spatial measurements from MaNGA in corresponding colors.
First of all, we find two SF-sBDEs (MaNGA 1-251788 and 1-258380), compared to the center of these galaxies, show slightly older age and lower $\ewha$ in the outer part.
At the same time, $\ewhd$ does not show significant changes, indicating that they both have star formation with a longer time scale.
And the outer region of MaNGA 1-51810 extends toward both larger $\ewha$ and $\ewhd$.
This is consistent with our results that these SF-sBDEs can be regarded as strong rejuvenated systems with longer star formation timescales.
For Q-sBDEs, it is worth noting that we don't obtain enough points of $\ewhd$ with S/N$>$5 in MaNGA 1-295446.
MaNGA 1-415777 doesn't exhibit significant changes in all three diagrams, while MaNGA 1-176908 and 1-295446 show wide distributions in the $\dn$-$\ewha$ relation.
This suggests that they may experience patchy but low- or medium-level star formation in the outer regions of galaxies.
Furthermore, MaNGA 1-176908 exhibits a bimodal distribution in the $\ewha$-$\ewhd$ and $\ewhd$-$\dn$ relations.
This implies that rejuvenation events may also have been occurring within Q-sBDEs.

\section{Discussion}\label{sec:Discussion}
ETGs are often characterized as ``red and dead'' galaxies, with little or even no cold gas.
However, this paradigm is considered too simplistic, as the development of many $\hi$ observations, such as HIPASS \citep{Meyer2004}, ALFALFA \citep{Alighieri2007,Grossi2009}, and ATLAS$\rm ^{3D}$ \citep{Serra2012,Serra2014}.
At current observational depths, the $\hi$ detection rate of ETGs is about 20\%, with a significant dependence on both the environment and mass \citep{Grossi2009,Ashley2019}.
When compared to LTGs, ETGs do indeed tend to exhibit gas-poor properties, especially ellipticals.
In the morphological classification sample of SDSS DR7 from \citet{Dominguez2018}, with redshift less than 0.1, ellipticals and lenticulars account for 8.2\% and 33.1\%, respectively, of the total number of galaxies.
After matching with the FASHI-ALFALFA combined sample, the fractions drop to 1.0\% and 4.5\%, respectively.
It is worth emphasizing that ellipticals constitute only a small fraction in the $\hi$ sample, and their ratio to lenticulars is also lower than in the optical sample.
In this study, we focus on a rare population, the $\hi$-rich sBDEs only, and study their basic properties, to uncover the origin of their $\hi$ gas and star formation. 
Despite machine learning classifying them as ellipticals, we refer to them as sBDEs because we allow for the presence of low surface brightness features at the outskirts of galaxies, distinguishing them from normal ellipticals.

\subsection{Possible mechanisms to obtain cold gas in $\hi$-rich sBDEs}
The MZR and two FMRs of sBDEs can provide insights into the origin of the gas.
If gas is of internal origin, as the metallicity continuously increases due to star formation, galaxies are expected to be located above the MZR and FMRs.
However, we find that most $\hi$-rich sBDEs do not exhibit higher metallicity in the MZR, and more importantly, most of them show significantly lower metallicity in the FMR relations (Figure \ref{fig:MZR.pdf}). 
The environment of $\hi$-rich sBDEs also indicates that the abundant cold gas cannot be attributed solely to internal origin. Instead, it is the low-density and star-forming environmental conditions that are the key for $\hi$-rich sBDEs to acquire gas (Figure \ref{fig:environment_1.pdf} and \ref{fig:environment_2.pdf}).
Furthermore, we do indeed find several misaligned galaxies in our sample from the MaNGA survey (Figure \ref{fig:MaNGA_new.pdf}).

Overall, these results suggest that the primary mechanisms for sBDEs to obtain gas are statistically external mechanisms, such as gas accretion from the surrounding environment or wet mergers \citep{Lagos2014,Lagos2015,Griffith2019,Lee2023}.
However, internal mechanisms may also play a key role in some individual galaxies, such as the two galaxies with unusually high metallicity in both MZR and FMRs (Figure \ref{fig:MZR.pdf}), which would require further observational data and research.

\subsection{Rejuvenated star formation in $\hi$-rich SF-sBDEs}\label{subsec:trigger_for_sf}
Some physical mechanisms could efficiently consume cold gas, expel gas from galaxies, and prevent the accretion of gas from the environment.
Once galaxies deplete their cold gas, the high-density environment and AGNs feedback can play crucial roles in the continuous suppression of star formation \citep[e.g.,][]{Giovanelli1983,Grossi2009,Peng2020,Kurinchi-Vendhan2023}.
However, the sBDEs in our sample have sufficient gas, but most of them (70\%) are still classified as quiescent galaxies.
AGNs do not have a significant impact on our sample, and the emission lines of Q-sBDEs in our sample are primarily dominated by DIG (see Section \ref{subsec:Metallicity} and Figure \ref{fig:BPT.pdf}).
Furthermore, although the reservoir of gas in sBDEs can be attributed to the galaxies being in an $\hi$-rich environment favorable for gas accretion (see Figure \ref{fig:environment_1.pdf}), it does not seem to be a key factor in determining star formation efficiency (see Figure \ref{fig:environment_2.pdf}).

The distribution of gas in galaxies may play a crucial role.
We find that the stacked $\hi$ profiles of SF-sBDEs are narrower and more concentrated, while those of Q-sBDEs are wider and similar to the quiescent control sample (Figure \ref{fig:spectra.pdf}).
Radial transport of gas is necessary to sustain star formation in galaxies \citep{Krumholz2018}.
However, the $\hi$ gas in these Q-sBDEs may form a ring-like structure in the outer region of galaxies.
Their high angular momentum makes it difficult for them to flow into galaxies \citep{Renzini2018,Peng2020,Lu2022}.
Meanwhile, these low-density $\hi$ gas outside the galaxies may not reach the threshold for transformation to $\rm H_{2}$ gas, and it is very inefficient to convert gas into stars \citep{Bigiel2008, Shi2011}.

Combining our results with spatially resolved studies can help us better understand the properties of $\hi$-rich sBDEs.
\citet{Lee2023} used the data from Mapping Nearby Galaxies at APO (MaNGA) to investigate the origin of star formation in ETGs.
Our results of stacked $\hi$ spectra and gas-phase metallicity indicate that the distributions of gas and star formation differ from those of star-forming LTGs.
This is consistent with their findings that star-forming ETGs show more concentrated ionized gas and different age gradients when compared to star-forming LTGs.
Besides, \cite{Lee2023} also find that more star-forming ETGs are classified as misalignment or counter-rotators compared to other types of galaxies.
A misaligned stellar-gas disk could efficiently deplete the angular momentum of the external gas, driving gas inflow to the central region of galaxies and triggering star formation \citep{Bao2022}.
The external origins, such as wet mergers or gas accretion, are likely to form the misaligned disk. 
In this case, our $\hi$-rich sBDEs are perfect candidates for the misaligned stellar-gas galaxies, as shown by the MaNGA image in Figure 10.
Further observations using integral-field spectroscopy are needed in the future.
In our sBDEs sample, the absence of star-forming disk typically found in lenticulars makes it more difficult to transport the angular momentum of gas outwards \citep{EWang2022}, resulting in the presence of so many $\hi$-rich quiescent sBDEs.
We hypothesize that, under certain conditions such as the disk instabilities caused by mergers or interactions, the gas in these galaxies would be induced to radial inflow and have the potential to revitalize star formation.

\section{Summary}\label{sec:Summary}
In this work, we have selected 104 $\hi$-rich strongly bulge-dominated ETGs (sBDEs) from the FASHI-ALFALFA combined sample and classified them into star-forming and quiescent sBDEs based on their SFR and stellar mass.
We have analyzed their basic galaxy properties using some $\hi$ scaling relations and stellar population scaling relations based on $\ewha$, $\ewhd$, and $\dn$.
To better understand the origin of $\hi$ gas in sBDEs and the possible reason for star formation in SF-sBDEs, we have also analyzed their gas-phase metallicity, stacked $\hi$ spectra, environment, and spatially resolved MaNGA data.

Our main results are summarized as follows.
\begin{enumerate}
\item The SF-sBDEs and Q-sBDEs exhibit a lower gas fraction compared to normal $\hi$-rich galaxies in the FASHI-ALFALFA combined $\hi$ sample, where the latter typically represent the most late-type galaxies.
Furthermore, SF-sBDEs exhibit significantly higher SFE$_{\hi}$ and shorter gas depletion timescale ($\sim$ 3 Gyr) compared to normal $\hi$-rich galaxies.

\item The distribution of SF-sBDEs on the $\ewha$-$\ewhd$ diagram is comparable to that of normal star-forming galaxies, but statistically have slightly larger values of $\dn$.  
This is consistent with the idea that the SF-sBDEs are composited by an old stellar population as the normal ellipticals, as well as a young stellar population with an age of Gyr-timescale.  

\item  Most $\hi$-rich sBDEs shows lower gas-phase metallicity in MZR, $\rm{FMR_{SFR}}$ and $\rm{FMR_{\hi}}$ based on O3N2 calibrator, indicating the dilution of metallicity from the external metal-poor gas. 
When using N2O2 as the metallicity calibrator, these results are more significant because it is less influenced by DIG.

\item From the analysis of stacked $\hi$ spectra, SF-sBDEs show narrower, more concentrated, and single-peaked stacked profiles compared to the star-forming control sample, while the stacked profiles of Q-sBDEs are flatter compared to those of SF-sBDEs.

\item We have analyzed the large-scale environment using the statistical average galaxies number and quiescent fraction within certain projected physical distances for each sub-sample. The $\hi$-rich sBDEs are more commonly located in significantly lower-density environments and potentially have much more star-forming neighbors compared to typical ellipticals. 

\item Three (in six) sBDEs in the MaNGA survey show kinematic misaligned gas-star rotation. 
This supports the conclusion of an external origin for the gas in most of these galaxies. 

\end{enumerate}

Our results support that external mechanisms are the dominant sources for $\hi$-rich sBDEs.
SF-sBDEs are likely the strong rejuvenating system with abundant $\hi$ reservoir and high $\rm SFE_{\hi}$,  while the gas in Q-sBDEs may be located at the outskirts of galaxies,  resulting in very low star formation efficiency.
We plan to look for $\hi$-rich sBDEs in cosmological hydrodynamical simulations such as IllustrisTNG \citep{Nelson2019} to study their formation and evolution.

\section*{Acknowledgements}
This work is supported by the National Science Foundation of China (NSFC, Grant No. 12233008, 12125301, 12192220, 12192222, 12273037), the National Key R\&D Program of China (2023YFA1608100), the Strategic Priority Research Program of Chinese Academy of Sciences (Grant No. XDB 41000000), the China Manned Space Project (No. CMS-CSST-2021-A07), the Cyrus Chun Ying Tang Foundations, and the 111 Project for ``Observational and Theoretical Research on Dark Matter and Dark Energy'' (B23042). EW thanks support of the Start-up Fund of the University of Science and Technology of China (No. KY2030000200). Y.R. acknowledges supports from the CAS Pioneer Hundred Talents Program (Category B), as well as the USTC Research Funds of the Double First-Class Initiative (grant No.~YD2030002013)

Funding for the Sloan Digital Sky Survey IV has been provided by the Alfred P. Sloan Foundation, the U.S. Department of Energy Office of Science, and the Participating Institutions. SDSS- IV acknowledges support and resources from the Center for High-Performance Computing at the University of Utah. The SDSS web site is www.sdss.org.
SDSS-IV is managed by the Astrophysical Research Consortium for the Participating Institutions of the SDSS Collaboration including the Brazilian Participation Group, the Carnegie Institution for Science, Carnegie Mellon University, the Chilean Participation Group, the French Participation Group, Harvard-Smithsonian Center for Astrophysics, Instituto de Astrof\'isica de Canarias, The Johns Hopkins University, Kavli Institute for the Physics and Mathematics of the Universe (IPMU) / University of Tokyo, Lawrence Berkeley National Laboratory, Leibniz Institut f\"ur Astrophysik Potsdam (AIP), Max-Planck-Institut f\"ur Astronomie (MPIA Heidelberg), Max-Planck-Institut f\"ur Astrophysik (MPA Garching), Max-Planck-Institut f\"ur Extraterrestrische Physik (MPE), National Astronomical Observatory of China, New Mexico State University, New York University, University of Notre Dame, Observatório Nacional / MCTI, The Ohio State University, Pennsylvania State University, Shanghai Astronomical Observatory, United Kingdom Participation Group, Universidad Nacional Aut\'onoma de M\'exico, University of Arizona, University of Colorado Boulder, University of Oxford, University of Portsmouth, University of Utah, University of Virginia, University of Washington, University of Wisconsin, Vanderbilt University, and Yale University.
Funding for the NASA-Sloan Atlas has been provided by the NASA Astrophysics Data Analysis Program (08-ADP08-0072) and the NSF (AST-1211644).

\appendix
\section{The metallicity properties based on N2O2 calibration of $\hi$-rich sBDEs}\label{appendix:N2O2}
\begin{figure*}[htbp]
	\centering
	\includegraphics[width=0.9\linewidth]{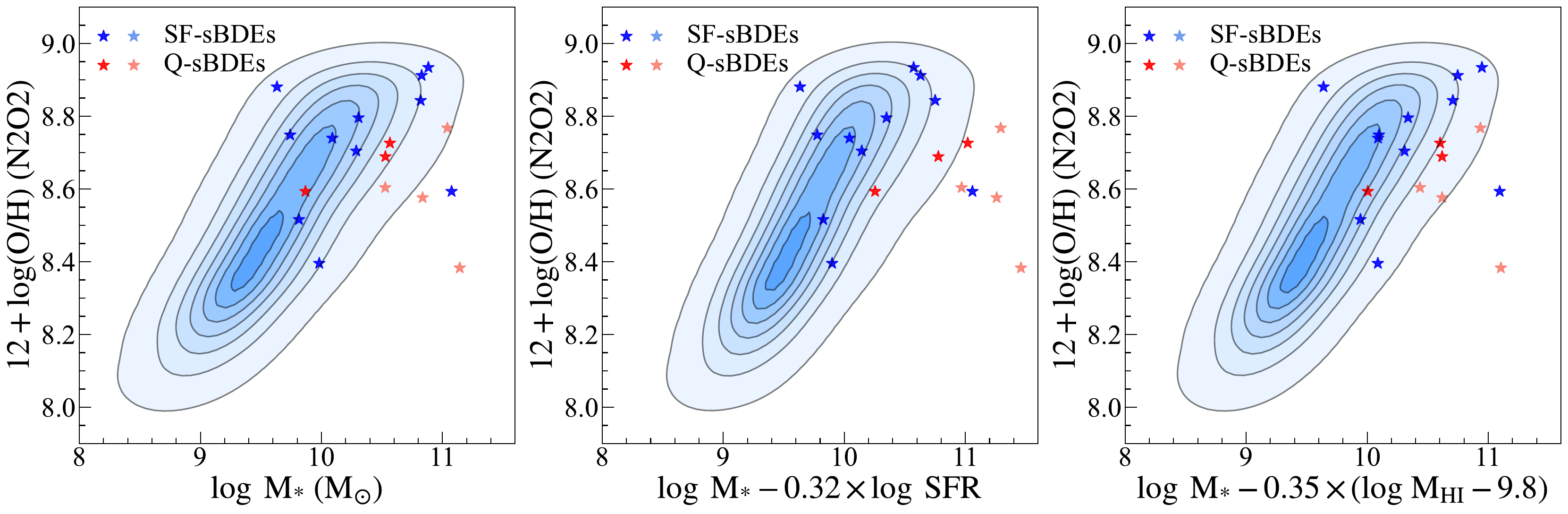}
	\caption{Same as Figure \ref{fig:MZR.pdf}, however using the N2O2 calibration from \citet{Sanders2018}.}
	\label{fig:MZR_N2O2.pdf}
\end{figure*}

In order to avoid contamination from DIG, we use the N2O2 calibrator from \citet{Sanders2018} to estimate the gas-phase metallicity \citep{Zhang2017,Kewley2019}.
Figure \ref{fig:MZR_N2O2.pdf} provides the mass-metallicity relation using the N2O2 system, but we only select 19 galaxies under the criteria of $\ewha$ $>$ 3 \AA\, and $-$1.3 $\rm < \log(N2O2)<$ 0.0.
We can find that $\hi$-rich sBDEs show significantly lower metallicity.

\startlongtable
\begin{deluxetable}{lccccccc}
\tablecaption{Galaxy Properties for sBDEs.}
\label{tab:data}
\centering
\tabcolsep=6pt
\renewcommand\arraystretch{1}
\tablehead{
\colhead{SDSS ObjID} &\colhead{Sample$^{a}$} & \colhead{RA (J2000)} & \colhead{DEC (J2000)}& \colhead{redshift} & \colhead{$\log$\, $\mstar$} & \colhead{$\log$\, $\rm SFR$} & \colhead{$\log$\,$\mhi$}\\
\colhead{} & \colhead{} & \colhead{(deg)} & \colhead{(deg)} & \colhead{} & \colhead{(M$_{\odot}$)}& \colhead{(M$_{\odot}$ yr$^{-1}$)}& \colhead{(M$_{\odot}$)}
}
\startdata
587730772817739967&ALFALFA&6.495&13.929&0.04196&10.7&-1.34&9.71\\
587724199351812182&ALFALFA&15.995&15.247&0.04174&10.81&0.73&10.08\\
587724197205835940&ALFALFA&19.445&13.323&0.04793&10.22&-1.19&10.12\\
587728906634592521&ALFALFA&114.032&30.215&0.02331&9.58&-0.32&9.22\\
587739153888837728&ALFALFA&117.195&16.646&0.05027&10.09&0.15&9.8\\
587739114696671327&ALFALFA&117.485&16.097&0.04085&10.67&-0.76&9.55\\
587731680111099933&ALFALFA&118.505&29.421&0.03609&10.38&-0.78&9.72\\
587732157389013064&ALFALFA&118.586&25.859&0.04164&10.5&0.35&9.73\\
587732157391306759&ALFALFA&122.823&29.509&0.01994&10.58&-1.61&9.89\\
587735236346707979&ALFALFA&124.994&26.386&0.01983&10.24&-0.53&9.56\\
587727944563622074&ALFALFA&133.559&2.241&0.05824&11.14&-1.06&10.12\\
587739116316786879&ALFALFA&137.096&28.269&0.04767&11.04&-0.79&10.1\\
587739158186754245&ALFALFA&138.827&28.315&0.04714&10.69&-1.01&9.95\\
588023046404636779&ALFALFA&139.805&19.702&0.02729&9.85&-1.08&9.47\\
587739377237622995&ALFALFA&140.004&26.739&0.04628&10.79&-1.22&9.76\\
587738409786671281&ALFALFA&146.054&11.146&0.03974&10.02&-1.28&9.74\\
587745245237805124&ALFALFA&147.771&17.056&0.02816&9.87&-1.21&9.41\\
587738616481185898&ALFALFA&149.721&31.62&0.02142&10.57&-1.41&9.61\\
587742567856472073&ALFALFA&149.799&16.478&0.02644&10.49&-1.41&9.66\\
587745402005422094&ALFALFA&150.209&13.737&0.03251&10.32&-1.57&9.83\\
587741532242444423&ALFALFA&155.197&26.112&0.03897&10.55&-1.31&9.76\\
587741816247812149&ALFALFA&155.407&21.687&0.03926&10.71&-1.33&9.78\\
587738411401478200&ALFALFA&155.514&13.769&0.01863&10.17&-1.51&9.53\\
588017702383648896&ALFALFA&156.796&10.025&0.03241&11.14&-0.99&9.91\\
587732703939461289&ALFALFA&159.805&7.858&0.03107&10.26&-1.46&9.23\\
587735349102182531&ALFALFA&161.762&14.263&0.03243&10.68&-1.38&9.74\\
588010359078912094&ALFALFA&163.269&4.493&0.04165&10.9&-1.26&10.18\\
587732578313175222&ALFALFA&167.82&7.336&0.04548&10.74&-1.26&10.2\\
587732771586375760&ALFALFA&168.481&9.643&0.04127&10.58&-0.73&9.95\\
587741709955039271&ALFALFA&170.72&27.584&0.03341&10.95&-1.15&10.0\\
588017567632261212&ALFALFA&173.223&13.164&0.03452&10.12&-1.18&9.7\\
587741727111315572&ALFALFA&174.223&25.452&0.04923&10.84&-1.32&10.42\\
588010880367525939&ALFALFA&175.023&5.865&0.02103&9.46&-2.0&9.11\\
587742773483143242&ALFALFA&175.743&16.488&0.03444&11.12&-0.67&9.79\\
587739646211588152&ALFALFA&176.404&31.3&0.00593&9.35&-0.41&8.78\\
587735347500417057&ALFALFA&182.565&14.327&0.04299&10.24&0.23&9.93\\
587738570850304033&ALFALFA&183.279&16.086&0.02599&10.61&-1.29&9.64\\
588017726012194878&ALFALFA&187.503&8.104&0.03737&-1.0&-0.51&9.46\\
587729157889917059&ALFALFA&188.458&3.944&0.04391&9.92&-1.1&9.79\\
587741726043865103&ALFALFA&190.106&24.904&0.04743&10.7&-0.68&9.92\\
587729158966542513&ALFALFA&195.05&4.922&0.0479&10.19&-1.15&9.99\\
587736541473931384&ALFALFA&197.935&9.11&0.04472&11.05&-0.99&10.05\\
587741721753157723&ALFALFA&202.811&26.27&0.04594&10.2&-0.74&9.79\\
588010879306432606&ALFALFA&204.051&4.74&0.03434&10.74&-0.41&9.93\\
587739504477667416&ALFALFA&205.547&29.896&0.02727&10.33&-1.66&9.77\\
587736542015127621&ALFALFA&208.002&8.882&0.03766&10.64&0.85&9.79\\
587739132416557077&ALFALFA&208.297&35.579&0.03978&10.77&0.19&10.19\\
587726015081480261&ALFALFA&208.736&2.096&0.01466&9.04&-0.48&9.14\\
588017724947759115&ALFALFA&208.87&6.596&0.02408&9.85&0.43&9.58\\
587739708480815182&ALFALFA&208.884&28.648&0.03462&10.67&-1.53&9.97\\
587729160046510114&ALFALFA&209.281&5.252&0.03968&10.86&0.84&9.75\\
587736542553899149&ALFALFA&212.341&8.907&0.0235&10.51&-1.15&9.78\\
588017991230292116&ALFALFA&212.901&8.81&0.02283&10.77&-1.38&9.11\\
587726101484404918&ALFALFA&215.814&4.302&0.05654&10.77&-0.77&10.05\\
587742629060804612&ALFALFA&220.301&14.89&0.03053&10.42&-1.01&10.0\\
587729158441402460&ALFALFA&221.855&3.441&0.02742&10.53&-1.38&10.06\\
587742578068750468&ALFALFA&223.103&16.847&0.04779&10.65&-0.63&10.04\\
587736543096340508&ALFALFA&225.195&8.144&0.04771&10.6&-1.06&10.07\\
587736478137319778&ALFALFA&238.874&7.597&0.04498&10.1&-1.23&9.98\\
588017990705545693&ALFALFA&240.621&5.154&0.03756&10.29&-0.82&9.85\\
587742611349438852&ALFALFA&242.067&10.109&0.01572&9.31&-1.08&8.77\\
587742627998531885&ALFALFA&246.276&8.382&0.03511&10.31&-0.13&9.71\\
587736920509251921&ALFALFA&247.428&25.304&0.04274&10.15&-1.1&10.04\\
587727223015342192&ALFALFA&339.208&14.387&0.01752&10.19&-1.3&9.17\\
587739845383356595&ALFALFA&217.575&21.747&0.01783&10.35&0.19&9.79\\
587736940378128696&ALFALFA&231.89&28.852&0.03201&10.22&-0.83&9.41\\
587742615635165195&ALFALFA&239.518&15.317&0.03757&10.82&0.22&10.12\\
587736813137035343&ALFALFA&241.208&8.481&0.01746&10.41&-0.68&9.41\\
587735744230916340&ALFALFA&249.366&26.205&0.05217&10.91&-1.28&10.1\\
587741491448250465&ALFALFA \& FASHI&175.016&30.593&0.03262&10.89&-1.33&10.17\\
587739458835054741&ALFALFA \& FASHI&219.441&28.059&0.04408&10.74&-0.99&10.07\\
587724242306203664&FASHI&54.905&-5.633&0.02091&10.15&-0.13&9.56\\
587725470662590755&FASHI&116.16&38.717&0.0631&10.92&0.35&9.75\\
587725551190802727&FASHI&113.439&36.679&0.01555&9.46&-0.36&9.61\\
587728932955291700&FASHI&138.301&51.281&0.06561&10.83&0.62&10.03\\
587729387155357807&FASHI&195.527&59.721&0.02729&10.13&-1.0&9.31\\
587729407545967137&FASHI&256.103&33.376&0.02964&10.67&-1.25&9.31\\
587732471463411876&FASHI&133.298&37.135&0.04979&10.88&0.97&9.62\\
587733440511410223&FASHI&237.067&42.931&0.03563&10.57&-1.42&9.7\\
587733605328093283&FASHI&210.019&58.37&0.02629&10.34&-1.24&8.96\\
587735429621743636&FASHI&212.307&46.242&0.03997&9.63&-0.01&9.78\\
587735490823913594&FASHI&229.161&41.03&0.03134&10.98&-1.16&8.94\\
587735662622670894&FASHI&148.994&37.156&0.01706&10.09&-1.57&9.44\\
587735665312399516&FASHI&232.087&46.187&0.03757&9.99&-1.17&9.61\\
587735743690244302&FASHI&241.368&31.526&0.05645&10.45&0.43&9.72\\
587738195043745901&FASHI&126.07&57.288&0.02625&10.64&-1.47&9.62\\
587738196651540694&FASHI&118.34&52.408&0.06366&10.8&-1.12&10.12\\
587738574068187225&FASHI&193.165&40.103&0.02624&10.76&-0.9&9.67\\
587738946146861087&FASHI&184.733&39.14&0.03694&10.98&-1.36&9.41\\
587739407320219766&FASHI&188.271&35.025&0.03219&9.98&0.26&9.5\\
587739407850668215&FASHI&170.462&34.628&0.0417&10.35&-1.51&9.48\\
587739406236188869&FASHI&160.366&31.703&0.03493&10.62&-1.1&9.5\\
587739721903374358&FASHI&199.985&30.119&0.02342&10.53&-0.78&9.54\\
588007004162293771&FASHI&127.872&48.626&0.05313&11.08&0.05&9.74\\
588013381666734111&FASHI&171.736&50.597&0.02321&10.59&-1.6&9.98\\
588017605209751562&FASHI&149.925&39.657&0.0285&10.49&-1.48&9.3\\
588017625628409940&FASHI&215.229&40.121&0.01754&9.99&-0.34&8.6\\
588017626691207193&FASHI&181.546&45.149&0.06654&10.97&1.19&10.26\\
588017721717555362&FASHI&181.072&42.995&0.07332&10.96&0.79&10.28\\
588017948277473344&FASHI&182.406&42.319&0.02256&9.74&-0.11&8.78\\
588017977289474168&FASHI&205.272&37.077&0.03435&10.29&-1.62&9.32\\
588017979438137462&FASHI&209.135&38.043&0.05013&10.29&0.45&9.74\\
588018254829977709&FASHI&236.425&40.212&0.02578&9.81&-0.03&9.43\\
588295842857287748&FASHI&170.059&50.025&0.02601&10.22&-1.49&8.64\\
\enddata
\end{deluxetable}

\begin{flushleft}
Notes. $^{a}$ The sources detected by both ALFALFA and FASHI sample used ALFALFA measurements.
\end{flushleft}

\bibliography{sBDEs_HI_accepted.bib}{}
\bibliographystyle{aasjournal}
\end{document}